\documentclass[twocolumn,twocolappendix,times,tighten]{aastex631}

\usepackage[normalem]{ulem}
\usepackage{bm}
\usepackage{color}

\newcommand*{\eqref}[1]{(\ref{#1})}

\newcommand*{\mt}{\mathrm}
\newcommand*{\unit}[1]{\;\mt{#1}}
\newcommand*{\abt}{\mathord{\sim}}
\newcommand*{\ptl}{\partial}
\newcommand*{\dtl}{\mathrm{d}}
\newcommand*{\del}{\nabla}

\renewcommand{\vec}[1]{\boldsymbol{#1}}
\newcommand*{\uvec}[1]{\hat{\boldsymbol{#1}}}

 \newcommand{\sgn}{\mathrm{sgn}}

\newcommand*{\dtlff}[2]{\frac{\mathrm{d} #1}{\mathrm{d} #2}}

\newcommand*{\bp}{\beta_{\mathrm{p}}}
\newcommand*{\Ma}{\mathcal{M}_{\mathrm{A}}}
\newcommand*{\Mms}{\mathcal{M}_{\mathrm{ms}}}
\newcommand*{\Ms}{\mathcal{M}_{\mathrm{s}}}
\newcommand*{\Mw}{\mathcal{M}_{\mathrm{w}}}
\newcommand*{\me}{m_{\mathrm{e}}}
\newcommand*{\mi}{m_{\mathrm{i}}}
\newcommand*{\mime}{m_{\mathrm{i}}/m_{\mathrm{e}}}
\newcommand*{\Omce}{\Omega_{\mathrm{ce}}}
\newcommand*{\Omci}{\Omega_{\mathrm{ci}}}
\newcommand*{\omp}{\omega_{\mathrm{pe}}}
\newcommand*{\ompe}{\omega_{\mathrm{pe}}}

\newcommand*{\de}{d_{\mathrm{e}}}
\newcommand*{\di}{d_{\mathrm{i}}}
\newcommand*{\prll}{\parallel}

\newcommand*{\Te}{T_{\mathrm{e}}}
\newcommand*{\TeTi}{T_{\mathrm{e}}/T_{\mathrm{i}}}
\newcommand*{\Ti}{T_{\mathrm{i}}}
\newcommand*{\kB}{k_{\mathrm{B}}}

\newcommand*{\delgame}{\Theta_\mathrm{e}}
\newcommand*{\thetaBn}{\theta_{Bn}}
\newcommand*{\thetaBk}{\theta_{Bk}}
\newcommand*{\thetakn}{\theta_{kn}}
\newcommand*{\debye}{\lambda_\mathrm{De}}

\newcommand*{\phidht}{\phi_\mathrm{HT}}
\newcommand*{\phiprll}{\phi_\parallel}
\newcommand*{\phigrid}{\phi_{\parallel,\mathrm{grid}}}
\newcommand*{\phiamb}{\phi_{\parallel,\mathrm{amb}}}

\newcommand*{\phinif}{\phi_\mathrm{NIF}}

\newcommand*{\vA}{v_{\mathrm{A}}}

\shorttitle{Oblique shock heating (\today)}
\shortauthors{Tran and Sironi (\today)}
\received{\today}

\begin{document}

\title{Electron Heating in 2D Particle-in-Cell Simulations of Quasi-Perpendicular Low-Beta
Shocks}

\correspondingauthor{Aaron Tran, Lorenzo Sironi}
\email{aaron.tran@columbia.edu, lsironi@astro.columbia.edu}

\author[0000-0003-3483-4890]{Aaron Tran}
\altaffiliation{Now at Department of Physics, University of Wisconsin--Madison, USA}
\affiliation{Department of Astronomy, Columbia University \\
538 W 120th St.~MC~5246, New York, NY 10027, USA}

\author[0000-0002-1227-2754]{Lorenzo Sironi}
\affiliation{Department of Astronomy, Columbia University \\
538 W 120th St.~MC~5246, New York, NY 10027, USA}

\begin{abstract}
We measure the thermal electron energization in 1D and 2D particle-in-cell
(PIC) simulations of quasi-perpendicular, low-beta ($\bp=0.25$) collisionless
ion-electron shocks with mass ratio $\mi/\me=200$, fast Mach number
$\Mms=1$--$4$, and upstream magnetic field angle $\thetaBn = 55$--$85^\circ$
from shock normal $\uvec{n}$.
It is known that shock electron heating is described by an ambipolar,
$\vec{B}$-parallel electric potential jump, $\Delta\phiprll$, that
scales roughly linearly
with the electron temperature jump.
Our simulations have
$\Delta\phiprll/(0.5 \mi {u_\mt{sh}}^2) \sim 0.1$--$0.2$ in units of the
pre-shock ions' bulk kinetic energy, in agreement with prior measurements and
simulations.
Different ways to measure $\phiprll$, including the use of de Hoffmann-Teller
frame fields, agree to tens-of-percent accuracy.
Neglecting off-diagonal electron pressure tensor terms can lead to a systematic
underestimate of $\phiprll$ in our low-$\bp$ shocks.
We further focus on two $\thetaBn=65^\circ$ shocks: a $\Ms=4$ ($\Ma=1.8$) case
with a long, $30\di$ precursor of whistler waves along $\uvec{n}$, and a
$\Ms=7$ ($\Ma=3.2$) case with a shorter, $5\di$ precursor of whistlers oblique
to both $\uvec{n}$ and $\vec{B}$; $\di$ is the ion skin depth.
Within the precursors, $\phiprll$ has a secular rise towards the
shock along multiple whistler wavelengths and also has localized
spikes within magnetic troughs.
In a 1D simulation of the $\Ms=4$, $\thetaBn=65^\circ$ case, $\phiprll$ shows a
weak dependence on the electron plasma-to-cyclotron frequency ratio
$\ompe/\Omce$, and $\phiprll$ decreases by a factor of $2$
as $\mi/\me$ is raised to the true
proton-electron value of $1836$.
\end{abstract}


\section{Introduction}

Shocks heat and compress their host medium.
In an ordinary fluid, collisions mediate heating; different particle species in
the fluid share the same temperature after passing through a shock.
In a low-density collisionless plasma, like the solar wind or
interstellar medium, shocks partition their energy between various plasma waves
and sub-populations of ions and electrons in a more complex way.
We wish to know how much energy goes to electrons, and how and where within the
shock the energy is gained, in order to
(1) provide sub-grid electron heating prescriptions that may be used in
two-fluid or hybrid fluid/kinetic simulations, and
(2) help interpret observations of heliospheric and astrophysical shocks.

In this study, we take Earth's quasi-perpendicular, low-beta magnetospheric bow
shock as an exemplar system.
Typical parameters are
upstream plasma beta (thermal/magnetic pressure ratio) $\bp \sim 0.1$--$10$,
upstream magnetic field angle from shock normal $\thetaBn \sim 55$--$90^\circ$,
and fast magnetosonic Mach number $\Mms \sim 1$--$10$
\citep[Figure~1]{farris1993}.
At such shocks, electrons may reflect from the shock front or leak from
downstream to upstream, whereas thermal ions are generally confined to the
downstream plasma after crossing the shock.

Electron heating can be described using a well-established cross-shock
potential model \citep{goodrich1984,scudder1995,hull1998}.
In a shock's de Hoffmann-Teller (HT) frame, electron and ion bulk flows are
parallel to $\vec{B}$ both upstream and downstream.
Entering the shock, a $\vec{B}$-parallel electric potential $\phiprll(x)$
varying along the shock-normal coordinate $x$ boosts electrons' parallel
velocities.
Small-scale scattering within the shock is expected to cool and smooth the
post-shock electron distribution as compared to the adiabatic Liouville-mapping
limit (i.e., test-particle evolution in static, macroscopic (ion-scale)
$\vec{B}(x)$ and $\phiprll(x)$)
\citep{scudder1986-iii,schwartz2014}.

The cross-shock potential model works and recovers trends in observed
post-shock temperature anisotropy \citep{scudder1986-iii,hull1998,lefebvre2007}.
But, the model is silent on the origin of small-scale scattering waves required
to create stable post-shock electron distributions \citep{schwartz2014}, e.g.,
flat-tops in Earth's magnetosheath \citep{feldman1983-earth}.
Scattering waves diffuse electron velocities and may alter electron pressure;
the electron pressure tensor divergence in Ohm's law sets $\phiprll$, which in
turn sets the free particle energy available to drive scattering waves
\citep{veltri1990,veltri1993-i,veltri1993-ii}.

Wave-particle interactions
also mediate electron heating, in a
complementary view.
Ion-scale whistler waves oblique to $\vec{B}$ are a prime suspect.
Low-Mach shocks host large-amplitude whistler precursors with a variety of
propagation angles
\citep{ramirez-velez2012,wilson2017,davis2021}.
Whistlers may radiate from the shock ramp
\citep{krasnoselskikh2002,dimmock2019}
or be generated by instabilities in the shock foot
\citep{wu1984,winske1985,muschietti2017}.
These whistlers have low-frequency $\vec{B}$-parallel electric fields that may trap and
heat electrons, and strong electron trapping (parallel electric field energy
comparable to the electrons' thermal energy) induces phase mixing and
short-wavelength electrostatic structures
\citep{mcbride1972,matsukiyo2003,matsukiyo2006,matsukiyo2010}.

The connection between scattering waves, electron heating, and the cross-shock
potential (or, ion-scale fields and waves more generally)
is not yet fully understood.
Recent Magnetospheric Multiscale (MMS) mission studies have given new insights.
For example, \citet{chen2018} suggest that the cross-shock potential could
arise from the cumulative effect of oblique, non-linear whistler wave
$\vec{E}$-fields; \citet{sun2022} suggest that strong, Debye-scale
double layers may help form the cross-shock potential.
\citet{cohen2019} have measured the electrostatic potential jump across an
interplanetary shock, along with collocated, strong, high-frequency
electrostatic fields.
\citet{hull2020} showed the onset of electron heating as a transition
from adiabatic to non-adiabatic electron distributions over multiple
wavelengths of a shock's oblique whistler precursor.
Much recent work has been done on electron scattering by coherent low- and
high-frequency whistlers at shocks
\citep{hull2012,wilson2016-accel,oka2017,oka2019,page2021,artemyev2022,shi2023},
including the construction and use of general frameworks such as stochastic
shock drift acceleration \citep{katou2019,amano2020} and magnetic pumping
\citep{lichko2020}, with particular success in explaining non-thermal
distributions.

In this manuscript, we use 1D and 2D particle-in-cell (PIC) simulations of
collisionless ion-electron shocks to
measure the strength and structure of the cross-shock potential.
We adopt $\bp=0.25$ ($\beta_\mt{e}=\beta_\mt{i}=0.125$), reasonable albeit on the low end
for Earth's bow shock \citep{farris1993}, which emphasizes the shock's
whistler-mode structure and reduces computational cost.
Our study builds upon prior fully-kinetic studies of quasi-perpendicular
shocks by
\citet{forslund1984,lembege1987,liewer1991,savoini1994,krauss-varban1995,
savoini1995,krasnoselskikh2002,scholer2007,riquelme2011,bohdan2022,morris2022},
and it complements recent simulations focused on perpendicular or
nearly-perpendicular ($\thetaBn \ge 80^\circ$) shocks
\citep{matsukiyo2003,scholer2003,scholer2006,matsukiyo2006,matsukiyo2012,
umeda2012-mtsi,yang2018,tran2020}.

We seek to address a few general questions.
\emph{First}, how much do electrons heat ($\Delta \Te$, $\TeTi$) across
quasi-perpendicular shocks?
Does heating occur at the ramp (steepest density rise), foot (region in which
shock-reflected ions execute a single Larmor gyration), or farther upstream
within a precursor?
How does the heating amount and location depend on shock parameters ($\Mms$,
$\thetaBn$) and simulation parameters (mass ratio, domain
dimensionality)?
\emph{Second}, what is the origin and nature of the cross-shock potential in
PIC simulations?
What structure does $\phiprll(x)$ possess along the shock-normal coordinate
$x$, beyond approximations such as $\phiprll(x) \propto B(x)$
\citep{hull2000-isee,hull2000-model,artemyev2022}?
How wide is the potential jump in $x$?
\emph{Third}, because $\phiprll$ is not easy to directly measure in the
heliosphere, how faithfully do various observational proxies (e.g., the
HT-frame electric field) reproduce $\phiprll$?

\section{Shock Setup}

\subsection{Simulation Methods}

We simulate 1D and 2D ion-electron shocks using the relativistic
particle-in-cell (PIC) code TRISTAN-MP \citep{buneman1993, spitkovsky2005}.
Our setup follows \citet{sironi2009, guo2014-accel, guo2017, tran2020}.
We use Gaussian CGS units throughout.

To form a shock, we inject upstream plasma with velocity $-u_0 \uvec{x}$ that
reflects off a conducting wall at $x=0$ in a uniform Cartesian
$(x,y,z)$ grid.
Coordinate unit vectors are $\uvec{x},\uvec{y},\uvec{z}$.
The reflected plasma interacts with upstream plasma to drive a shock
traveling towards $+\uvec{x}$.
The domain's left-side ($x=0$) boundary specularly reflects all particles in the
simulation frame ($v_x \to -v_x$).
The domain's right-side $x$ boundary expands ahead of the shock, continuously
adding new plasma to the upstream; the expansion speed is manually chosen to
out-run waves, thermal electrons leaking from downstream to upstream,
and mirror-reflected (shock-drift accelerated) particles that may all travel
ahead of the shock.
Particles and waves exiting the right-side $x$ boundary are deleted.
The $y$ and $z$ boundaries are periodic, and we simulate 2D shocks in the
$x$-$y$ plane.
All three Cartesian components of particle velocities and electromagnetic
fields are tracked in both 1D and 2D shocks.

The injected upstream plasma comprises Maxwellian ions and electrons with equal
density $n_0$ and temperature $T_0$, carrying a magnetic field $\vec{B}$ with
magnitude $B_0$ and angle $\thetaBn$ with respect to shock normal
($\uvec{n} = \uvec{x}$).
The upstream $\vec{B}$ lies in the $x$-$y$ plane.
The ion and electron plasma frequencies
$\omega_\mt{p\{i,e\}} = \sqrt{4\pi n_0 e^2/m_\mt{\{i,e\}}}$,
inertial lengths $d_\mt{\{i,e\}} = c/\omega_\mt{p\{i,e\}}$,
and cyclotron frequencies
$\Omega_\mt{c\{i,e\}} = e B_0/(m_\mt{\{i,e\}}c)$
are defined with respect to upstream plasma properties.
Subscripts $\mt{i}$ and $\mt{e}$ denote ions and electrons respectively.
Subscripts $\perp$ and $\prll$ are generally defined with respect to
local $\vec{B}$.

\begin{figure}
    \includegraphics[width=3.375in]{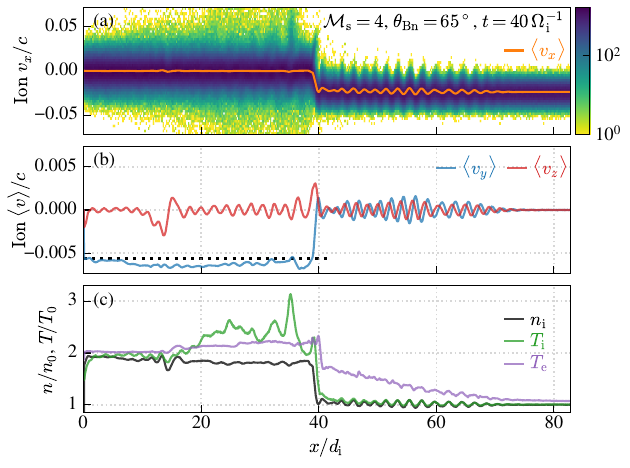}
    \caption{
        Structure of a 2D $\Ms=4$, $\thetaBn=65^\circ$ shock along the
        shock-normal coordinate $x$, averaged over the transverse $y$
        coordinate.
        (a) Ion $x-v_x$ distribution with overlaid mean $\langle v_x \rangle$
        (orange).
        (b) Ion bulk velocities $\langle v_y \rangle$ (blue) and
        $\langle v_z \rangle$ (red).
        The ion $\langle v_y \rangle$ velocity deflects downstream of the
        shock, $0 < x < 40\di$,  following the MHD Rankine-Hugoniot prediction
        (dotted black line).
        (c) Ion density $n_\mt{i}$ (black), ion temperature $T_\mt{i}$ (green),
        and electron temperature $T_\mt{e}$ (purple) normalized to upstream
        values $n_0$, $T_0$.
    }
    \label{fig:v3iy}
\end{figure}

A shock is specified by several dimensionless parameters:
$\Ms$, (or $\Ma$, $\Mms$), $\bp$, $\thetaBn$, $\Gamma$, and $c_\mt{s}/c$.
The total plasma beta $\bp = 16\pi n_0 \kB T_0 /{B_0}^2$.
The fast magnetosonic, sonic, and Alfv\'{e}n Mach numbers are
$\Mms = u_\mt{sh} / \sqrt{{c_\mt{s}}^2 + {v_\mt{A}}^2}$,
$\Ms = u_\mt{sh} / c_\mt{s}$, and
$\Ma = u_\mt{sh} / v_\mt{A}$
respectively.\footnote{
    The Mach number $\Mms$ is defined with respect to the MHD fast speed
    for propagation perpendicular to $\vec{B}$.
}
Here, $u_\mt{sh}$ is upstream plasma speed in the shock's rest frame.
The sound speed $c_\mt{s} = \sqrt{2\Gamma\kB T_0/(\mi+\me)}$
and the Alfv\'{e}n speed $v_\mt{A} = B_0/\sqrt{4\pi n_0(\mi+\me)}$,
with $\Gamma$ the one-fluid adiabatic index.
We scale all velocities with respect to $c$ by choosing any of the ratios
$c_\mt{s}/c \propto v_\mt{A}/c \propto (\omp/\Omce)^{-1}
\propto \sqrt{\kB T_0/(\me c^2)}$,
for fixed $\{\Ms, \bp, \mime\}$.

How do we set $u_0$ to obtain a desired Mach number?
In the simulation frame, the downstream (post-shock) plasma has bulk $u_x = 0$,
and the shock speed is $u_\mt{sh}/r$, where $r$ is the post-shock compression
ratio.  We relate $u_0$ and $u_\mt{sh}$ by an explicit boost:
\[
    u_0 = \frac{u_\mt{sh} - u_\mt{sh}/r}{1 - ({u_\mt{sh}}^2/r)/c^2}
\]
taking both $u_0, u_\mt{sh} > 0$.
To compute $r = r(\Ms, \Gamma, \bp, \thetaBn)$ from the MHD Rankine-Hugoniot
(R-H) conditions \citep{tidman1971}, we must guess the effective one-fluid
adiabatic index $\Gamma$, which is not known \emph{a priori} for our
kinetic plasma.
We adopt $\Gamma=5/3$ to set $u_0$ for all simulations in this manuscript.
We expect $\Gamma\approx 5/3$ for stronger or more oblique shocks with good
coupling between ion velocities parallel and perpendicular to $\vec{B}$.
But $\Gamma\approx 2$, corresponding to a 2D non-relativistic gas, may be more
appropriate for weak, nearly perpendicular shocks with little ion
isotropization.
Our choice of $\Gamma=5/3$ to initialize shocks that are best described by
$\Gamma \approx 2$ can cause our stated Mach numbers to be inaccurate by
$\abt 10\%$ or less.

We define the shock position $x_\mt{sh}$ to be located where the magnetic
fluctuation $B/B_0 - 1$ declines to one-half its maximum value, measured
rightwards (increasing $x$) from the maximum value of $B/B_0$ just after the
shock ramp.

In oblique MHD shocks (i.e., $\thetaBn \ne 90^\circ$) with $\vec{B}$ in the
$x$-$y$ plane, upstream plasma with bulk velocity $u_y = 0$
acquire a transverse drift $u_y \ne 0$ after transiting the shock.
Our left-side ($x=0$) boundary accommodates this drift by imposing
$E_z = (u_y/c) B_0 \cos\thetaBn$ in the 10 left-most $x$-cells of the Yee mesh.
Particles reflect 5 $x$-cells rightwards of the imposed-field region.
The boundary $u_y$ is computed from the MHD R-H jump conditions, and it is
typically $\abt 10\%$ of $u_\mt{sh}$ \citep[Figures~1.5--1.6]{tidman1971}.
The boundary mimics a conducting wall that slides in the $y$ direction.

Our choice of particle reflection procedure at the left boundary can alter
post-shock electron distributions.
Specular reflection, $v_x \to -v_x$, mixes field-parallel and perpendicular
particle momenta in the downstream rest frame and thus may alter particle
distributions' anisotropy.
Other choices are possible, e.g., \citet{krauss-varban1995}
let particles escape the domain at both upstream and downstream domain
boundaries, with new particles continuously injected from Maxwellian
distributions.
In nature, the post-shock boundary condition may be set by global system
effects;
e.g., \citet{mitchell2012,mitchell2013,mitchell2014,schwartz2019,horaites2021}
construct collisionless magnetosheath models wherein electron distributions are
regulated by the two points at which magnetosheath $\vec{B}$ field lines
penetrate the bow shock.
Because we do not consider global transport effects upon shock structure,
our simulations can only inform us about electron heating local to a shock,
under the assumption that post-shock electrons catching up to the shock (i.e.,
traveling from downstream to upstream) have a similar distribution as those
streaming away from the shock.

Figure~\ref{fig:v3iy}(a) shows the structure of an example 2D, $\Ms=4$,
$\thetaBn=65^\circ$ shock.
Incoming ions with $\langle v_x \rangle = -u_0 = -0.0238 c$ were reflected at
$x=0$ to form a shock now at $x \sim 40 \;\di$, traveling from left to right.
The ion $v_x$ distribution oscillates ahead of the shock ($x\gtrsim 40\;\di$)
as part of a precursor whistler wave train.
Figure~\ref{fig:v3iy}(b) shows the bulk ion velocities $\langle v_y \rangle$
and $\langle v_z \rangle$ for our example shock; the black dotted line shows
the transverse $u_y$ deflection predicted by the MHD R-H conditions and also
used to set the left-side $x=0$ wall's transverse velocity.
The transverse $\langle v_y \rangle$ drift is nearly constant throughout the
post-shock region ($0 < x < 40\di$).
Figure~\ref{fig:v3iy}(c) shows the shock structure in ion density and
ion/electron temperatures.
The electron temperature rises ahead of the shock within the same region
($x \sim 40$ to $70 \;\di$) as the ion bulk-velocity oscillations.

\subsection{Shock Parameters} \label{sec:shock-param}

We fix $\mime=200$ and $\bp=0.25$;
we vary the magnetic obliquity $\thetaBn = 55$--$85^\circ$
and sonic Mach number $\Ms = 3$--$10$.
The transverse shock width is $5.4 \; \di$ for 2D runs.
We choose the upstream electron temperature $\delgame = \kB T_0 /(\me c^2)$ to
keep post-shock electrons marginally non-relativistic; lower $\delgame$ is more
realistic and costlier.
For $\Ms=3$--$4$ we take $\delgame = 0.01$;
for $\Ms=5$, $\delgame = 0.007$;
for $\Ms=7$, $\delgame = 0.005$;
for $\Ms=10$, $\delgame = 0.003$.
Choosing $\delgame$ also sets
$\ompe/\Omce = \sqrt{\bp/(4\delgame)} = 2.5$--$4.5$.
The solar wind is
much colder, with $\delgame \sim 2\times10^{-5}$ and
$\ompe/\Omce \sim 50$--$100$; most PIC simulations do not attain realistic
$\delgame$ and $\ompe/\Omce$
due to computing cost \citep{wilson2021}.
Table~\ref{tab:param} in the Appendices provides more simulation
parameters.

The grid spacing $\Delta x = \Delta y = 0.1 \de$ and the timestep
$\Delta t = 0.045 {\omp}^{-1}$, so $c = 0.45 \Delta x / \Delta t$.
We use $128$ particles per cell for 2D runs, and $2048$ for 1D runs.
The upstream thermal electron gyroradius
$r_\mt{Le} = \left(\Gamma \bp/4\right)^{1/2} \de$
is resolved with $\abt 3$ cells at $\bp=0.25$.
The upstream Debye length $\debye = \delgame^{1/2} \de$
is marginally resolved with one cell for $\Ms=3$--$5$ and not resolved for
$\Ms=7$--$10$.
At each step in the PIC algorithm, the electric current is smoothed with $N=32$
sweeps of a 3-point (``1-2-1'') digital filter in each axis \citep[Appendix
C]{birdsall1991}; the filter has a half-power cut-off at
$k \approx \sqrt{2/N} (\Delta x)^{-1} = 2.5 {\de}^{-1}$.

\begin{figure}
    \includegraphics[width=3.375in]{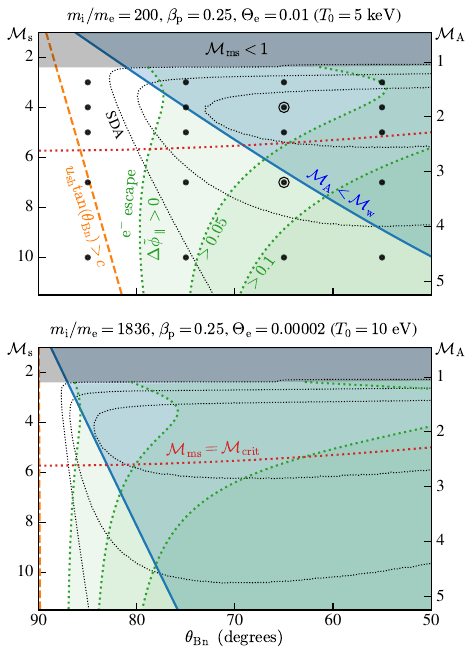}
    \caption{
        Regimes of $\bp=0.25$ shock behavior in $\Ms$--$\thetaBn$ space for
        our simulations (top panel) and true mass ratio $\mime=1836$ with solar
        wind-like $T_0 = 10\unit{eV}$ (bottom panel).
        Black dots in top panel mark our simulations; circled black dots
        are studied in more depth in Section~\ref{sec:case-study}.
        Black shaded region is $\Mms < 1$, calculating MHD fast
        speed with $\thetaBn$ dependence (unlike definition and use of $\Mms$
        elsewhere in manuscript; definitions agree at $\thetaBn=90^\circ$).
        Blue shaded region is whistler sub-critical, $\Ma < \Mw$.
        Green shaded region shows which shocks may permit thermal electrons to
        escape from downstream to upstream, for varying $\Delta\tilde{\phi}_\prll=0$,
        $0.05$, and $0.1$ (Equation~\eqref{eq:phinorm}).
        Red dotted line marks where $\Mms$ equals the critical Mach number
        \citep{marshall1955,kennel1985}.
        Orange dashed line is sub/super-luminal boundary for $\delgame=0.01$;
        note that the $\Ms=7,10$ and $\thetaBn=85^\circ$ simulations are
        marginally sub-luminal, not super-luminal, because of their lower
        $\delgame$ values.
        Black dotted contours bound the regions where the SDA reflection
        efficiency is $>1\%$ for varying $\Delta\tilde{\phi}_\prll = 0$, $0.05$, and
        $0.1$ (larger to smaller regions).
    }
    \label{fig:regimes}
\end{figure}

\section{Regime Map of Shock Parameters}

\begin{figure*}
    \includegraphics[width=\textwidth]{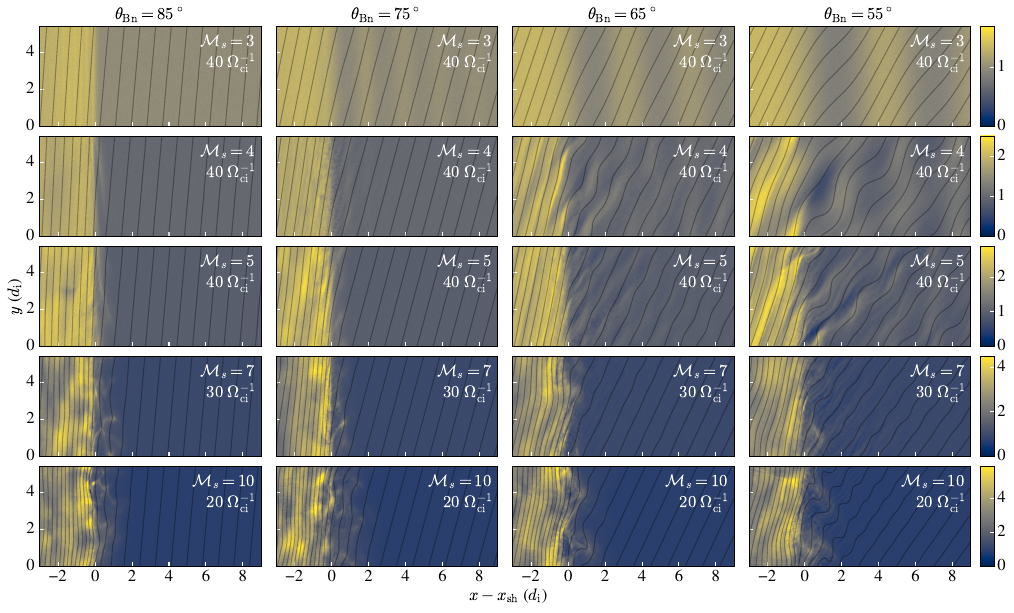}
    \caption{
        Ion density at end of each 2D shock simulation, with ending time $t=20$
        to $40\;{\Omci}^{-1}$ labeled in each panel.
        Columns vary magnetic obliquity from $\thetaBn=85^{\circ}$ (left) to
        $55^{\circ}$ (right);
        rows vary shock strength from $\Ms=3$ (top) to $10$ (bottom).
        Faint black lines trace magnetic field lines.
        See online journal for animated time evolution from $t=0$ to end of
        all simulations.
    }
    \label{fig:mosaicdens}
\end{figure*}

\begin{figure*}
    \centering
    \includegraphics[width=\textwidth]{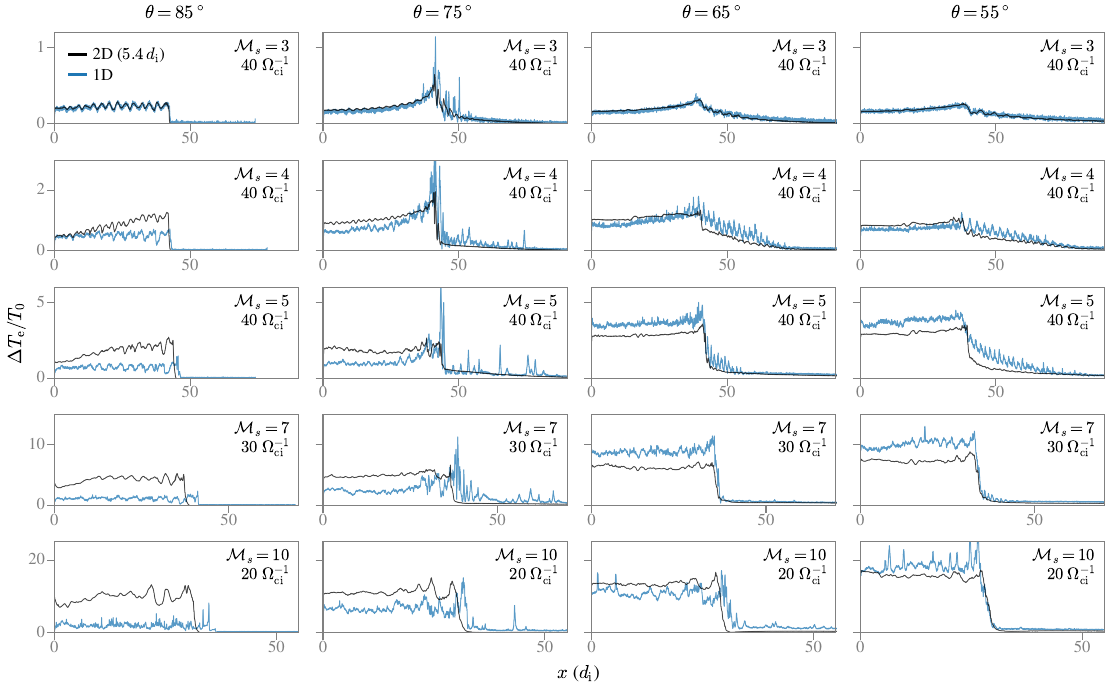}
    \caption{
        Electron temperature $\Te$ along the shock-normal coordinate $x$ for
        shock simulations of varying magnetic obliquity $\thetaBn = 85^\circ$
        (left column) to $55^\circ$ (right column), and varying Mach number
        $\Ms = 3$ (top row) to $10$ (bottom row).
        Temperature is a density-weighted average along $y$.
        Comparing 1D (blue curves) versus 2D (black curves) simulations, we see
        that nearly perpendicular, super-luminal or marginally sub-luminal
        shocks with $\thetaBn = 85^\circ$ to $75^\circ$ require 2D geometry for
        substantial heating.
        Sub-luminal shocks with $\thetaBn = 65^\circ$--$55^\circ$ heat
        electrons to a similar level in 1D and 2D, within a factor of two.
        Some 2D shocks heat electrons less than in 1D
        (i.e., $\Ms=5$--$7$ and $\thetaBn=65$--$55^\circ$).
    }
    \label{fig:mosaicte}
\end{figure*}

Our simulations inhabit various regimes of collisionless shock behavior.
Figure~\ref{fig:regimes} shows approximately where precursors of leaking
thermal electrons (green shaded), waves (blue shaded), and shock-reflected
electrons (convex, black dotted contours) may appear upstream of a shock as a
function of $\Ms$ and $\thetaBn$.
The top panel is for our simulated shock parameters.
The bottom panel is for more realistic, solar wind-like parameters; adopting
the true proton/electron mass ratio $\mime=1836$, in particular, permits
thermal electron and whistler wave precursors for a large region of
($\Ms,\thetaBn$) parameter space.
Figure~\ref{fig:regimes} (top panel) serves as an interpretive key to the
presence or absence of shock precursors in Figures~\ref{fig:mosaicdens} and
\ref{fig:mosaicte}.
Let us explore each piece of Figure~\ref{fig:regimes} in turn.

First, some notation.
We rescale the parallel potential by the incoming ion bulk kinetic energy:
\begin{equation} \label{eq:phinorm}
    \tilde{\phi}_\prll
    = \frac{e \phi_\prll}{\frac{1}{2} \mi {u_\mt{sh}}^2}
    \, .
\end{equation}
The tilde ($\sim$) will mean the same normalization for other electric
potentials throughout this manuscript.
And, the prefix $\Delta$ (e.g., $\Delta\tilde{\phi}_\prll$) will generally mean
a quantity's cross-shock jump value, except for the grid step sizes $\Delta
x$, $\Delta y$, and $\Delta t$.

In the blue-shaded region of Figure~\ref{fig:regimes}, whistlers propagating
along shock normal have phase and group speeds faster than the shock speed, so
whistler wave trains may ``phase stand'' ahead of the shock
\citep{tidman1968,kennel1985,krasnoselskikh2002,oka2006}.
Such a precursor requires $\Ma < \mathcal{M}_\mt{w}$, where the whistler Mach
number
\[
    \mathcal{M}_\mt{w}
    = \frac{v_\mt{w}}{v_\mt{A}}
    = \frac{1}{2} \sqrt{\frac{\mi}{\me}} \cos\thetaBn \; ,
\]
and $v_\mt{w}$ is the maximum phase speed of an oblique whistler based on an
approximate cold-plasma dispersion relation \citep{krasnoselskikh2002}.

The orange dashed line of Figure~\ref{fig:regimes} denotes the
sub/super-luminal boundary.
Shocks are subluminal for $u_\mt{sh} \tan\thetaBn < c$, where $\thetaBn$ is
measured in the shock frame (also called normal incidence frame, NIF);  we use
$\thetaBn$ measured in the simulation (downstream rest) frame, since $\vec{B}$
is nearly frame invariant when the relative velocity between NIF and simulation
frame is non-relativistic.
Shocks right of the boundary are sub-luminal; left, super-luminal.
All sub-luminal shocks may be boosted into the de-Hoffmann Teller (HT) frame.
All super-luminal shocks may be boosted into a perpendicular shock frame
wherein the shock is stationary and both upstream and downstream $\vec{B}$
fields are oriented $90^\circ$ from shock normal $\uvec{n}$
\citep{drury1983,kirk1989,begelman1990}.
For smaller $\delgame$, ceteris paribus, the super-/sub-luminal boundary shifts
leftward so that super-luminal shocks occupy a smaller region of the plot,
while the other curves and regions in the regime plot are unchanged.
Our $\Ms=5$, $\Ms=7$, and $\Ms=10$ simulations use smaller values of
$\delgame$ than shown in Figure~\ref{fig:regimes} (Table~\ref{tab:param}).
The two simulations with $\Ms=7,10$ and $\thetaBn=85^\circ$ are both marginally
sub-luminal.

In the green-shaded regions of Figure~\ref{fig:regimes}, downstream thermal
electrons may escape from the shock into the upstream.
For a given ($\thetaBn$, $\Ms$, $\Delta\tilde{\phi}_\prll$), we define a thermal escape
criterion as follows.
Consider a thermal electron with downstream rest frame four-velocity
$\gamma\beta_\prll$ equal to the mean value for a Maxwell-J\"{u}ttner
distribution of temperature $T_{\mt{e}2}$, and $\gamma\beta_\perp = 0$ (i.e.,
zero pitch angle).
Here $\gamma$ is the Lorentz factor; $\beta$ is three-velocity in units of
$c$.
We compute an estimated post-shock electron temperature $T_{\mt{e}2}$ from the
non-relativistic oblique MHD R-H conditions
\citep{tidman1971} assuming post-shock equipartition,
$T_{\mt{e}2} = T_{\mt{i}2}$.
We choose the sign of $\gamma\beta_\prll > 0$ to indicate travel towards
$+\uvec{x}$
in the downstream rest frame.
Now, boost into the HT frame.
If the electron has HT-frame $\beta_\prll > 0$ and HT-frame kinetic energy
exceeding the cross-shock potential jump, i.e.,
\[
    (\gamma - 1) \me c^2 > \Delta\phiprll
    \, ,
\]
then we conclude that ``typical'' thermal electrons may escape from downstream
to upstream.
Equivalently, for non-relativistic HT-frame boosts, the escape criterion is
approximately equal to
\[
    \frac{1}{2} \me \left( c_{se2} \cos\theta_{Bn2} - \frac{u_\mt{sh}}{r} \right)^2
    > \Delta\phiprll
    \, ,
\]
where $c_{se2}$ is the downstream electron thermal speed interpreted here as a
$\vec{B}$-parallel velocity, and $\theta_{Bn2}$ is the post-shock magnetic
field angle with respect to shock normal.
Figure~\ref{fig:regimes} shows regions where electrons may escape for
$\Delta\tilde{\phi}_\prll = 0, 0.05, 0.1$, with successively darker green shading
corresponding to larger $\Delta\tilde{\phi}_\prll$.
Our assumptions somewhat overestimate the possibility of post-shock thermal
electron escape; the quasi-perpendicular shocks in this manuscript typically
have post-shock $T_{\mt{e}2} < T_{\mt{i}2}$ and $\Delta\tilde{\phi}_\prll \sim 0.1$--$0.2$.
Our escape criterion also neglects mirroring as electrons travel from high to
low $\vec{B}$ magnitude, which would aid the escape of electrons with non-zero
pitch angle as perpendicular energy transfers to parallel energy.

The red dotted line of Figure~\ref{fig:regimes} denotes the critical Mach
number \citep{marshall1955,edmiston1984,kennel1985}; for $\Mms$ exceeding this
Mach number,
ion reflection is expected to become important for shock dissipation.

The black dotted contours in Figure~\ref{fig:regimes} denote where shock drift
acceleration (SDA) may reflect $>1\%$ of incoming thermal electrons.
Successive (smaller) contours correspond to $\Delta\tilde{\phi}_\prll = 0$,
$0.05$, and $0.1$.
We calculate the SDA reflection efficiency by applying
\citet[Equations~(13)--(14)]{guo2014-accel} to a Monte-Carlo sampling of upstream
Maxwell-J\"{u}ttner distributions for a grid of shock parameters
$(\Ms,\thetaBn)$.
The reflection efficiency is defined as the ratio between the number of
particles with upstream HT-frame $v_\prll<0$ (approaching the shock) that would
mirror and reflect upon reaching the shock front, compared to the total number
of particles with upstream HT-frame $v_\prll<0$.

\section{Overview of Shock Simulations}

Figure~\ref{fig:mosaicdens} shows the ion density structure
for our ($\Ms$, $\thetaBn$) parameter sweep of 2D shocks.
Shocks change from laminar to turbulent as the Mach number rises.
All $\Ms=3$ shocks appear laminar.
As $\thetaBn$ decreases from $85^\circ$ for the weaker shocks ($\Ms=3$--$5$),
shock precursors appear as coherent waves and
filaments at $x>x_\mt{sh}$ (Figure~\ref{fig:mosaicdens}), and also
spatially-coincident rises in $\Te$ (Figure~\ref{fig:mosaicte}).
Towards higher $\Ms$, ion reflection becomes important; we observe discrete
clumps of ions and distorted $\vec{B}$-field lines within
$2 \di$ of the shock ramp; this reflection is seen for all $\thetaBn$.
To limit the scope of this work, we cannot firmly identify all the
instabilities or linear modes in Figure~\ref{fig:mosaicdens}, but let us
speculate briefly.
Ion/electron beam drifts coupling to the whistler mode (various forms of
modified two-stream instability, aka MTSI)
\citep{wu1984,hellinger1999,matsukiyo2003,matsukiyo2006,muschietti2017}
could be responsible for fine compressive filamentation with wavevector nearly
perpendicular to $\vec{B}$ seen in the $\Ms=5$, $\thetaBn=65,55^\circ$
cases.
At $\Ms=10$ and $\thetaBn=55^\circ$, the localized field-line bending and less
coherent fluctuations (i.e., not clearly wave-like) may be a weaker form of the
turbulence and reconnection in 2D and 3D PIC quasi-parallel shocks simulated by
\citet{bessho2020,bessho2022,ng2022}.
At $\Ms=7$--$10$ and $\thetaBn\sim85$--$75^\circ$, the reflected ion clumps and
shock ramp fluctuations are likely similar to those reported by
\citet{tran2020}.

Figure~\ref{fig:mosaicte} shows the 1D $y$-averaged electron temperature $\Te$
for both 1D and 2D shocks.
The temperature in units of $m_\mt{e}c^2$ is the trace of
$\int \gamma \beta_{i} \beta_{j} f_\mt{e}(\vec{r},\vec{p}) d^3\vec{p}$
for electron distribution function $f_\mt{e}$ and tensor indices $i$,$j$.
In the near-perpendicular $\thetaBn=85^\circ$ and $75^\circ$ shocks,
electron heating is much greater in 2D than in 1D; a 2D domain is required for
non-adiabatic heating to occur within the shock ramp.
Ions' effective $\Gamma$ and hence the shock propagation speed both change
when comparing 1D and 2D, especially at $\Ms \gtrsim 5$.
Towards lower $\thetaBn$, the trend changes.
The weakest, oblique shocks generally have similar 1D and 2D heating (e.g.
$\Ms=3$, $\thetaBn=55^\circ$).
Some stronger oblique 2D shocks heat electrons less than their 1D counterparts
($\Ms=5$--$7$, $\thetaBn=65$--$55^\circ$).

\begin{figure}
    \includegraphics[width=3.375in]{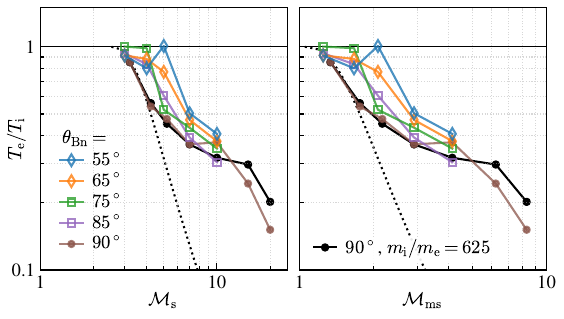}
    \caption{
        Post-shock electron-ion temperature ratio $\TeTi$ from our 2D runs as a
        function of sonic Mach number $\Ms$ (left) and magnetosonic Mach number
        $\Mms$ (right).
        The upstream plasma beta $\bp=0.25$ for all points.
        Dotted black line is baseline from adiabatic $\Gamma=2$ compression
        of electrons, without other ion-electron energy exchange,
        based on MHD R-H shock jump prediction at $\thetaBn=90^\circ$.
    }
    \label{fig:machscale}
\end{figure}

Figure~\ref{fig:machscale} plots the downstream electron/ion temperature ratio
$\TeTi$ for the 2D runs.
We measure $\TeTi$ in the downstream plasma within $x-x_\mt{sh}=-18$ to
$-3\di$; details are given in Appendix~\ref{app:deltapot-measure}.
For comparison, we also show exactly-perpendicular shock data at $\mime=200$
and $625$ that were previously presented in \citet{tran2020}.
The dotted black line corresponds to an adiabatic $\Gamma=2$ compression for
electrons based on the MHD R-H shock jump prediction
at $\thetaBn=90^\circ$.
We observe that the electron-ion temperature ratio at $\thetaBn=85^\circ$ is
similar to that at $\thetaBn=90^\circ$.
And, $\TeTi$ increases towards unity as $\thetaBn$ decreases.

\section{Parallel Potential Praxis}
\label{sec:phiprll}

In quasi-perpendicular shocks where electrons are well-magnetized, and the de
Hoffmann-Teller (HT) frame is accessible (i.e., the shock is subluminal), the
space physics community has an established description of electron heating.
Electron motion,
at 0th order,
is assumed to obey quasi-static, macroscopic fields
$\vec{B}(x)$, $\vec{E}(x)$ varying only along $x$ with no time dependence
\citep{feldman1983-earth,goodrich1984,scudder1986-iii,hull1998,hull2000-isee,hull2001}.
By ``macroscopic'', we mean that $\vec{B}$ and $\vec{E}$ are assumed to vary on
ion length scales.
Individual electron trajectories within the HT frame may be modeled as
conserving two invariants, energy $\varepsilon$ and magnetic moment $\mu$
\citep{hull1998}:
\begin{eqnarray}
    \varepsilon &= \frac{1}{2} m_e \left[{v_\perp}^2(x) + {v_\prll}^2(x)\right]
    - e \phi_\prll(x) \; , \label{eq:eom-ener} \\
    \mu &= \frac{m_e {v_\perp}^2(x)}{2 B(x)} \; , \label{eq:eom-mu}
\end{eqnarray}
where $\phi_\prll$ is an electrostatic potential that forms within the shock
transition due to the $\vec{B}$-parallel electric field component, $E_\prll$.
The parallel potential reads:
\begin{equation} \label{eq:phiprll}
    \phi_\prll(x)
    = \int_{x}^{\infty} \frac{E_\prll(x)}{\uvec{x}\cdot\uvec{b}(x)} \dtl x \; ,
\end{equation}
where $\uvec{b}$ is the magnetic field unit vector, and $E_\prll$ is set by the
electron momentum equation (i.e., generalized Ohm's law):
\[
    \vec{E} = - \frac{\vec{V}_\mt{e} \times \vec{B}}{c}
              - \frac{\del \cdot \vec{P}_\mt{e}}{e n_\mt{e}}
              - \frac{m_\mt{e}}{e} \frac{\dtl \vec{V}_\mt{e}}{\dtl t} \; .
\]
Here, $\vec{V}_\mt{e}$ is the electron bulk velocity,
$\vec{P}_\mt{e}$ is the electron pressure tensor, $n_\mt{e}$ is the electron
density, and
$\dtl/\dtl t = \ptl/\ptl t + \vec{V}_e \cdot\del$ is the total derivative.
Collisions are neglected.
The inertial term $\dtl\vec{V}_\mt{e}/\dtl t$ is small,\footnote{
    Take $eE_\mt{amb} \approx \Delta P_\mt{e}/(n_\mt{e} L)$
    and $eE_\mt{inertial} \approx m_\mt{e} (\Delta V_\mt{e}) V_\mt{e}/L$; here
    $\Delta P_\mt{e}$, $\Delta V_\mt{e}$ are the shock jumps in electron
    pressure and bulk velocity, and $L$ is the gradient length.
    Suppose $\Delta P_\mt{e} \sim P_\mt{e}$ and $\Delta V_\mt{e} \sim V_\mt{e}$.
    Then $E_\mt{amb}/E_\mt{inertial} \sim \Ms^{-2} (\mi/\me) \sim 10$ in our
    simulations and is larger in nature.
}
so the ambipolar term $\del \cdot \vec{P}_\mt{e}$ sets the parallel electric
field in all frames:
\[
    E_\prll
    \approx \uvec{b}\cdot\vec{E}_\mt{amb}
    = \frac{-\uvec{b}\cdot (\del \cdot \vec{P}_\mt{e})}{e n_\mt{e}} \, .
\]
In the HT frame, $\vec{V}_\mt{e}$ is parallel to $\vec{B}$, up to
usually-small correction terms localized within the shock layer
\citep{scudder1987}.
So, $\vec{E}_\mt{HT} \approx \vec{E}_\mt{amb}$ as well.
In the following discussion, we treat simulation-frame measurements of
$E_\parallel$, $\phi_\prll$, and $\vec{B}$ as equal to their HT-frame
values, because all three quantities are frame invariant to order
$\mathcal{O}(u/c)$ in boost velocity $u$.

This quasi-static, macroscopic description is not complete, for shocks host
high-frequency and short-wavelength fields that will scatter electrons and
break the invariants $\mu$ and $\varepsilon$
\citep{veltri1990,hull1998,see2013,schwartz2014}.
Within the context of this macroscopic-field model, scattering is required to
obtain physical (stable) electron distributions, and scattering will alter
$\vec{P}_\mt{e}$ and therefore $\phi_\prll$ as well.
Nevertheless, the macroscopic fields do successfully describe the electron
dynamics observed by spacecraft \citep{scudder1986-iii,lefebvre2007}.

Let us now compare various ways to measure $\phi_\prll$, which we summarize
in Figures~\ref{fig:liouville-fit-procedure} and \ref{fig:dc-proxies} for one
example shock simulation.
See also \citet[Section~3]{schwartz2021} for a recent, similar discussion
of $\phi_\prll$ measurement methods applied to MMS data.

\subsection{Liouville Mapping Measurement of the Parallel Potential}
\label{sec:phiprll-liouville}

We can estimate $\Delta\phiprll$ using Liouville mapping fits of the electron
distribution function, as is commonly done to interpret satellite data
\citep{scudder1986-iii,schwartz1988,hull1998,lefebvre2007,johlander2023}.
If electrons evolve adiabatically across a shock,
their distribution at small pitch angle will inflate in energy
following Liouville's theorem;
by comparing upstream and downstream electron distributions, the amount of
energy gain can be fitted to infer $\Delta\phiprll$.

We extract both upstream (unshocked) and downstream (shocked) electron
distribution function cuts along $p_\prll$ by selecting particles
with pitch angle $< 15^\circ$, all in the HT frame.
Downstream distributions are sampled from $x-x_\mt{sh} = -18$ to $-3 \di$.
Upstream distributions are sampled from
$x-x_\mt{sh}=65\di$ to $105\di$ for $\thetaBn=55^\circ$ shocks,
$45\di$ to $85 \di$ for $\thetaBn=65^\circ$ shocks, and
$25\di$ to $65 \di$ for $\thetaBn=75^\circ$ shocks.

\begin{figure}
    \includegraphics[width=3.375in]{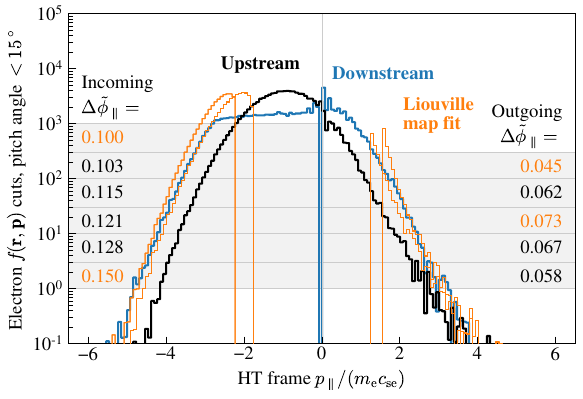}
    \caption{
        Liouville mapping: by comparing upstream (black curve) and downstream
        (blue curve) electron distribution cuts along the $p_\prll$ axis,
        we fit Liouville-mapped upstream distributions (orange curves) to the
        downstream distributions in order to estimate
        $\Delta\tilde{\phi}_\prll$ for different electron populations
        (incoming $p_\prll<0$ versus outgoing $p_\prll>0$) and for different
        bands in $f$.
        For incoming $p_\prll < 0$, the smallest and largest best-fit values
        of $\Delta\tilde{\phi}_\prll = 0.100$ and $0.150$
        are highlighted in
        orange alongside their corresponding Liouville-mapped distributions.
        For outgoing $p_\prll > 0$, the smallest and largest best-fit values
        of $\Delta\tilde{\phi}_\prll = 0.045$ and $0.073$
        are similarly highlighted.
        Momentum is scaled using the upstream electron thermal speed
        $c_{se} = \sqrt{\Gamma \kB T_0/\me}$.
        The shock is the same 2D, $\Ms=4$, $\thetaBn=65^\circ$ case in
        Figure~\ref{fig:dc-proxies}.
    }
    \label{fig:liouville-fit-procedure}
\end{figure}

Next, we map the sampled upstream particles to a notional
downstream state, specified by a magnetic compression ratio $B_2/B_0$ and
potential jump $\Delta\phiprll$.
We fix $B_2/B_0$ to its volume-averaged value in the region
$x-x_\mt{sh} = -18$ to $-3 \di$, while $\Delta\phiprll$ is a free parameter.
Let $u_\perp = \gamma\beta_\perp$ and $u_\prll = \gamma\beta_\prll$ be the
dimensionless four-velocity components of each particle.
The Liouville-mapped particle momenta $u_{\perp 2}$ and $u_{\prll 2}$,
written in terms of the upstream momenta $u_{\perp 0}$ and $u_{\prll 0}$,
are:
\begin{eqnarray}
    \label{eq:liouville-map-perp}
        u_{\perp 2} &= u_{\perp 0} \sqrt{B_2/B_0} \\
    \label{eq:liouville-map-gam}
        \gamma_2 &= \sqrt{1 + u_{\prll 0}^2 + u_{\perp 0}^2}
            + e\Delta\phiprll/(\me c^2) \\
    \label{eq:liouville-map-prll}
        u_{\prll 2} &= \sqrt{\gamma_2^2 - 1 - u_{\perp 2}^2}
        \, .
\end{eqnarray}
The mapping applies to both incoming ($u_{\prll 0} < 0$) and outgoing
($u_{\prll 0} > 0$) electron populations, so the outgoing electrons are
mapped backwards in time.
We then select particles with pitch angle $<15^\circ$ and bin in $p_\prll$ to
form a Liouville-mapped distribution (Figure \ref{fig:liouville-fit-procedure},
orange curves), which can be compared to the true downstream distribution
(Figure \ref{fig:liouville-fit-procedure}, blue curve).

The Liouville-mapped distributions are normalized to both momentum- and
real-space volumes.
The momentum-space volume is a cone fixed by the pitch-angle selection.
The real-space volume changes across the shock due to
$\vec{B}$-perpendicular compression and
$\vec{B}$-parallel velocity change across the shock; the ratio of 3D real-space volume
elements is
\begin{equation} \label{eq:dxdx0}
    \frac{\dtl \vec{x}_2}{\dtl \vec{x}_0}
    =
    \frac{B_0}{B_2}
    \frac{\beta_{\prll 2}}{\beta_{\prll 0}}
    \, .
\end{equation}
In the non-relativistic limit,
\[
    \frac{\beta_{\prll 2}}{\beta_{\prll 0}}
    = \left[
        1
        - \frac{2 e \Delta\phiprll}{\me c^2 \beta_{\prll 2}^2}
        + \left(\frac{\beta_{\perp 2}}{\beta_{\prll 2}}\right)^2
        \left(1 - \frac{B_0}{B_2}\right)
    \right]^{-1/2}
    \, .
\]
If two particles are separated along a $\vec{B}$ field line as they cross
the potential jump $\Delta \phiprll$, their time-staggered accelerations
will tend to increase their real-space separation.
This $\Delta\phiprll$-induced dilation is most important for small pitch angles
and small initial parallel velocities, i.e., $\beta_{\prll 0} \sim 0$.

In contrast to our particle-based procedure, some studies
\citep[e.g.,][]{chen2018,cohen2019} perform Liouville mapping by shifting the
distribution $f$ in energy coordinates,
$f(\varepsilon) \to f(\varepsilon + e\Delta\phiprll)$.
We find general agreement between the two procedures in one-off experiments
with our simulation data.

The Liouville-mapped distributions are fitted to the downstream
distributions using a least-squares fit that minimizes a cost function
\begin{equation} \label{eq:cost}
    \frac{1}{N} \sum_{i=1}^N
    \left[
    \log\left(
        \frac{ f_\mt{downstream}(p_{\prll,i}) }
             { f_\mt{mapped}(p_{\prll,i}) }
    \right)\right]^2
    \, ,
\end{equation}
where $N$ is the number of momentum bins.
We separately fit the $p_\prll<0$ and $p_\prll >0$ parts of $f$ within
multiple bands of $f$ values, manually chosen to lie below the downstream
flat-top $f$ value and above counting noise at high energies.

Figure~\ref{fig:liouville-fit-procedure} shows the Liouville mapping
procedure for the 2D, $\Ms=4$, $\thetaBn=65^\circ$ shock.
The mean best-fit
$\Delta\tilde{\phi}_\prll = 0.12$
for incoming electrons
(HT-frame $p_\prll < 0$, upstream to downstream).
The mean best-fit
$\Delta\tilde{\phi}_\prll = 0.061$
for outgoing electrons
(HT-frame $p_\prll > 0$, downstream to upstream).
Fit values have a $\pm \abt20\%$ range for incoming electrons and a larger
range for outgoing electrons.
Appendix~\ref{app:liouville} shows the Liouville-mapped distributions for
all the 2D shocks in our parameter sweep.

\subsection{Direct Measurement of the Parallel Potential}
\label{sec:phiprllmeth}

In PIC simulations, we can measure $\phiprll$ from global particle and
field information that is not always available in satellite measurements.
We consider only 1D, $y$-averaged measures of $\phi_\prll$ and
neglect the possibility of different potential jumps along different $\vec{B}$
field lines in 2D.
For this section of the manuscript, angle brackets $\langle\cdots\rangle$
denote transverse ($y$-axis) averaging.
Integrals have the same limits as in Equation~\eqref{eq:phiprll}.

First, we may measure $E_\parallel$ directly from the PIC mesh
(Figure~\ref{fig:dc-proxies}(b)):
\begin{equation} \label{eq:phigrid}
    \phi_{\prll,\mt{grid}}
    = \int \frac{\langle E_\parallel \rangle}{\langle \uvec{x}\cdot\uvec{b} \rangle}
    \;\dtl x \, .
\end{equation}

Second, we may measure $E_\parallel$ from the electron pressure tensor
divergence (Figure~\ref{fig:dc-proxies}(c)):
\begin{equation} \label{eq:phiamb}
    \phi_{\prll,\mt{amb}}
    = \int \left\langle
        \frac{-\uvec{b}\cdot\left(\del\cdot\vec{P}_\mt{e}\right)}{e n_\mt{e}}
        \right\rangle
        \frac{1}{\langle\uvec{x}\cdot\uvec{b}\rangle}
    \dtl x \, .
\end{equation}
We compute $\vec{P}_\mt{e}$ on a grid by depositing particles with a $5$ cell
flat-top kernel ($5\times5$ in 2D), and we evaluate $\del$ using a 2nd-order
centered finite difference.

Third, we can simplify Equation~\eqref{eq:phiamb}.  Assume a gyrotropic pressure
tensor
$\vec{P}_e = P_{e\prll} \uvec{b}\uvec{b} + P_{e\perp} (\vec{I} - \uvec{b}\uvec{b})$,
following \citet{goodrich1984}, to obtain
(Figure~\ref{fig:dc-proxies}(d)):
\begin{equation} \label{eq:phigyr}
    \phi_{\prll,\mt{gyr}} = \int \left\langle
        \frac{-1}{e n_\mt{e}}
        \left[
            \frac{\dtl P_{e\prll}}{\dtl x}
            - \left(P_{e\prll} - P_{e\perp}\right)
            \frac{\dtl \ln B}{\dtl x}
        \right]
    \right\rangle \dtl x \, .
\end{equation}
All $\dtl/\dtl y$ and $\dtl/\dtl z$ terms are neglected.
Note that the spatial average is taken over the entire integrand, unlike
Equations~\eqref{eq:phigrid} and \eqref{eq:phiamb}.

We have chosen a particular order for the transverse-average operations,
$\langle \uvec{b}\cdot\vec{E} \rangle / \langle \uvec{x}\cdot\uvec{b} \rangle$
for Equation~\eqref{eq:phigrid} and the same with $\vec{E}_\mt{amb}$ replacing
$\vec{E}$ for Equation \eqref{eq:phiamb}.
We could have instead averaged prior to all the vector projections, e.g.,
integrate
$\langle\uvec{b}\rangle \cdot \langle\vec{E}\rangle / \langle
\uvec{x}\cdot\uvec{b} \rangle$ in Equation~\eqref{eq:phigrid} (``early''
averaging).
Or, we could have deferred averaging until the entire integrand is formed,
e.g., integrate
$\langle (\uvec{b} \cdot \vec{E}) / (\uvec{x}\cdot\uvec{b}) \rangle$
in Equation~\eqref{eq:phigrid} (``late'' averaging).
We compared the three averaging procedures across the shock parameter range.
The ``late'' averaging gives poor results; it may be too sensitive to
local regions where $\uvec{x}\cdot\uvec{b}$ approaches zero, whereas
$\langle\uvec{x}\cdot\uvec{b}\rangle$ is constrained to be non-zero in the
other procedures due to $B_x$ conservation across the shock.
The ``early'' averaging agrees well with our adopted procedure in some cases,
but not all.
There is not a clear reason to favor or disfavor our adopted procedure as
compared to ``early'' averaging, due to the inherent approximation of
describing a 2D shock with a 1D $y$-averaged profile;
disagreement between the two procedures is due solely to 2D effects.

\subsection{Indirect Measurement of the Parallel Potential}
\label{sec:phiprllmethdht}

\begin{figure}
    \includegraphics[width=3.375in]{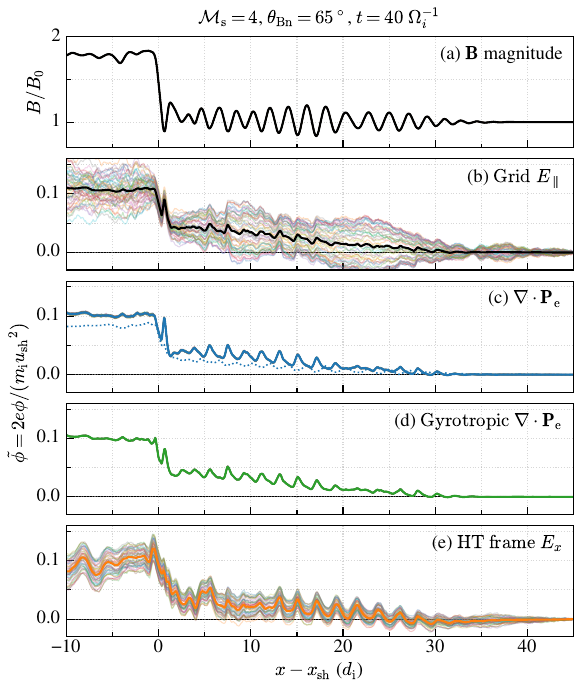}
    \caption{
        Parallel electrostatic potential measured four different ways in an
        example 2D, $\Ms=4$, $\thetaBn=65^\circ$ shock, using 50 evenly-spaced
        snapshots within $t=40.026$ to $40.123{\Omci}^{-1}$.
        (a) Magnetic field magnitude scaled to its upstream value, $B_0$.
        (b) Potential $\phi_{\prll,\mt{grid}}$ measured from PIC grid $E_\prll$.
        (c) Potential $\phi_{\prll,\mt{amb}}$ measured from electron pressure
        tensor divergence, $\del \cdot \vec{P}_\mt{e}$.
        Dotted line is the contribution from $\uvec{x}\cdot(\del \cdot \vec{P}_\mt{e})$
        (Equation~\eqref{eq:phiambx}), which samples only the shock-normal
        component of $\vec{E}_\mt{amb}$, to show that the non-$\uvec{x}$
        components of $\vec{E}_\mt{amb}$ contribute measurably to
        $\phi_{\prll,\mt{amb}}$.
        (d) Potential $\phi_{\prll,\mt{gyr}}$ measured from electron pressure
        assuming gyrotropy (Equation~\eqref{eq:phigyr}).
        (e) Potential $\phidht$ measured from the HT-frame electric field
        by integrating the shock-normal component $E_{\mt{HT},x}$.
        In all panels, faint colored lines are single time snapshots; thick
        lines are averages.
        Multiple time snapshots appear in panels (c)--(d), but they agree well
        enough so as to be indistinguishable.
    }
    \label{fig:dc-proxies}
\end{figure}

The HT-frame cross-shock potential (Figure~\ref{fig:dc-proxies}(e)),
\begin{equation} \label{eq:phidht}
    \phidht(x) = \int \vec{E}_\mt{HT}(x) \cdot \uvec{x} \,\dtl x
    \, ,
\end{equation}
will approximately equal the parallel potential $\phi_\prll(x)$ if
$\vec{E}_\mt{amb}$ points towards $+\uvec{x}$ in the HT frame
\citep{goodrich1984}.
The HT potential has been used to interpret electron heating in satellite
measurements, and it has been shown to agree with other proxies for
electron heating (most notably, Liouville mapping), within the large
systematic uncertainties of measuring low-frequency $\vec{E}$ fields in space
\citep{cohen2019,schwartz2021}.
We compute $\vec{E}_\mt{HT}$ by an explicit boost of simulation-frame fields with
a constant global boost velocity;
this may be contrasted with adaptive methods that account for local velocity
and $\vec{B}$ variations \citep{comisel2015,marghitu2017}.

Why should $\vec{E}_\mt{amb}$ point towards $+\uvec{x}$?
Suppose that $\vec{P}_\mt{e}$ is isotropic (scalar) and that the shock
structure varies solely along the shock-normal coordinate $x$.
Then, the ambipolar field
\[
    \vec{E}_\mt{amb}
    = - \frac{1}{e n_\mt{e}} \dtlff{P_\mt{e}}{x} \uvec{x}
    \, ,
\]
and the parallel potential
\[
    \phi_\prll
    = - \int \frac{1}{e n_\mt{e}} \dtlff{P_\mt{e}}{x}
    \, \dtl x
    \, .
\]
Since $\vec{E}_\mt{HT} \approx \vec{E}_\mt{amb}$,
then $\phidht \approx \phi_\prll$.

Next, let us consider a situation in which $\vec{E}_\mt{amb}$ may not point
towards $+\uvec{x}$.
Suppose that $\vec{P}_\mt{e}$ is gyrotropic and not isotropic.
Then, $\vec{P}_\mt{e}$ has off-diagonal terms proportional to
$(P_{\mt{e}\prll}-P_{\mt{e}\perp})$ because $\uvec{b}$ does not coincide with
a coordinate axis in the HT frame.
Let us still assume a shock varying solely along $x$; i.e., neglect $\dtl/\dtl
y$ and $\dtl/\dtl z$ terms.
Then, the ambipolar field
\[
    \vec{E}_\mt{amb}
    = - \frac{1}{e n_\mt{e}} \left(
        \dtlff{P_{\mt{e},xx}}{x} \uvec{x}
        + \dtlff{P_{\mt{e},xy}}{x} \uvec{y}
        + \dtlff{P_{\mt{e},xz}}{x} \uvec{z}
    \right)
    \, ,
\]
and the parallel potential
\[
    \phi_\prll
    = - \int \frac{1}{e n_\mt{e}} \left(
        \dtlff{P_{\mt{e},xx}}{x}
        + \frac{b_y}{b_x} \dtlff{P_{\mt{e},xy}}{x}
        + \frac{b_z}{b_x} \dtlff{P_{\mt{e},xz}}{x}
    \right)
    \, \dtl x
    \, .
\]
Here, $b_x$, $b_y$, $b_z$ are components of $\uvec{b}$;
notice $b_x = \uvec{x}\cdot\uvec{b}$.
Since $\vec{E}_\mt{HT} \approx \vec{E}_\mt{amb}$,
\begin{equation} \label{eq:phidht-pxx}
    \phidht
    = - \int \frac{1}{e n_\mt{e}}
        \dtlff{P_{\mt{e},xx}}{x}
    \, \dtl x \, .
\end{equation}
We see $\phi_\prll \ne \phidht$ due to the non-$x$ components of
$\vec{E}_\mt{amb}$ in this situation.\footnote{
    Analogous to the disagreement between normal-incidence and HT frame
    cross-shock potentials explained by \citet{goodrich1984}.
}
This electron anisotropy effect is just one of several higher-order corrections
to the quasi-static field model as discussed by \citet{scudder1987}.

\begin{figure*}
    \includegraphics[width=\textwidth]{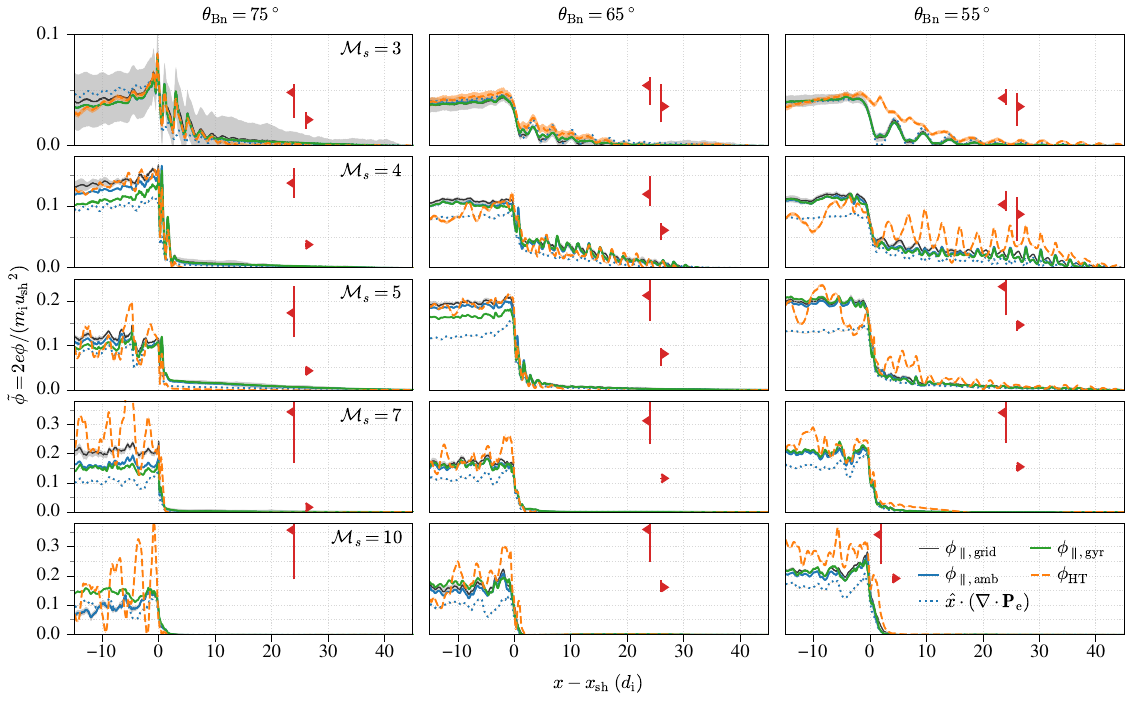}
    \caption{
        Parallel electrostatic potential measured in different ways for
        2D shocks.
        Black solid line is $\phi_{\prll,\mt{grid}}$, Equation~\eqref{eq:phigrid}.
        Blue solid line is $\phi_{\prll,\mt{amb}}$, Equation~\eqref{eq:phiamb}.
        Dotted blue line is same as Equation~\eqref{eq:phiamb}, except the integrand
        is replaced with $\langle\uvec{x}\cdot(\del\cdot\vec{P}_\mt{e})/(e n_\mt{e})\rangle$.
        Green solid line is $\phi_{\prll,\mt{gyr}}$, Equation~\eqref{eq:phigyr}.
        Orange dashed line is $\phidht$, Equation~\eqref{eq:phidht}.
        Shaded regions (gray, orange) represent uncertainty in the grid-based
        proxies $\phi_{\prll,\mt{grid}}$ and  $\phidht$; region widths are
        the expected standard error of the mean, $\pm \sigma/\sqrt{N}$ with
        $N=50$ for the number of time snapshots and $\sigma$ the standard
        deviation, computed at each $x$ position.
        Red triangles are mean of best-fit $\Delta\tilde{\phi}_\prll$ from
        Liouville mapping; left-facing triangle is from incoming (HT-frame
        $v_\prll < 0$) electrons, and right-facing triangle is from
        outgoing (HT-frame $v_\prll > 0$) electrons.
        Red vertical bars show the range of $\Delta\tilde{\phi}_\prll$ values
        inferred from different $f$-band fits of the same electron
        populations; see Figure~\ref{fig:liouville-fit-procedure} and
        Section~\ref{sec:phiprll-liouville}.
    }
    \label{fig:dc-proxies-all}
\end{figure*}

We test the importance of the off-diagonal $\vec{P}_\mt{e}$ terms in the
simulation coordinate system by computing
\begin{equation} \label{eq:phiambx}
    \int \left\langle
        \frac{-\uvec{x} \cdot (\del\cdot\vec{P}_\mt{e})}{e n_\mt{e}}
    \right\rangle
    \dtl x \, ,
\end{equation}
shown as a blue dotted line in Figures~\ref{fig:dc-proxies}(c) and
\ref{fig:dc-proxies-all}.

If $\vec{E}_\mt{HT} = \vec{E}_\mt{amb}$ and
the shock is 1D-like
($\dtl/\dtl y$ and $\dtl/\dtl z$ terms negligible), then
Equation~\eqref{eq:phiambx} resolves to Equation~\eqref{eq:phidht-pxx}, which
should equal $\phidht$.
If in addition $\vec{E}_\mt{amb}$ points towards $+\uvec{x}$,
then Equation~\eqref{eq:phiambx} should equal both $\phidht$ and $\phiamb$.
We shall see that these assumptions do not fully hold in our simulations.

\subsection{Parallel Potential Praxis -- Results}
\label{sec:phiprllres}

Figure~\ref{fig:dc-proxies} compares different proxies for $\phi_\prll$ in our
example $\Ms=4$, $\thetaBn=65^\circ$ shock (same as Figure~\ref{fig:v3iy}),
All potentials are scaled to the shock-frame upstream ion bulk kinetic energy
(Equation~\eqref{eq:phinorm}).
We use 50 snapshots with close time spacing to smooth out short-timescale
variation.
Individual snapshots are plotted as faint colored lines, while their average is
plotted as a thicker solid line.
The time spacing between snapshots is $\Delta x / \sqrt{\kB T_0/m_\mt{e}}$,
where $\sqrt{\kB T_0/m_\mt{e}}$ is an upstream electron thermal velocity;
the snapshots span $t=40.026$ to $40.123{\Omci}^{-1}$.
Our choice of time spacing allows thermal electrons to translate by $\gtrsim 1$
grid cell between snapshots, in order to decorrelate thermal fluctuations in
consecutive snapshots while keeping the shock stationary on ion scales.
We set $\phi=0$ at $x-x_\mt{sh}=45 d_\mt{i}$ for each $\phi_\prll$ proxy and
for each snapshot before averaging the data together.

What do we learn from Figure~\ref{fig:dc-proxies}?
We observe an extended rise in the potential ahead of the shock, spanning tens
of $\di$, corresponding to a precursor region of increased $\Te$ as seen in
Figure~\ref{fig:mosaicte}.
Local spikes in $\phi_\prll$ (potential wells for electrons) appear at the
position of magnetic troughs in the precursor wave train; a strong potential
spike appears in the magnetic trough immediately adjacent to the shock ramp at
$x=x_\mt{sh}$ (Figure~\ref{fig:dc-proxies}(a)).
The $E_\prll$ measured from the PIC mesh is noisy; multiple time snapshots must
be averaged to obtain a good measurement of $\phi_{\prll,\mt{grid}}$.
Both PIC field-based estimators $\phi_{\prll,\mt{grid}}$ and $\phidht$ show
more fluctuation than the particle-based estimators $\phi_{\prll,\mt{amb}}$ and
$\phi_{\prll,\mt{gyr}}$.
The particle-based estimators can give a good estimate of the potential
$\phi_\prll$ with a single snapshot.
Both particle-based estimators are in reasonable agreement with each other,
suggesting that gyrotropy is a good assumption.
The particle-based estimators are close to, albeit $\abt 5\%$ lower than,
$\phigrid$.

The net cross-shock jump in $\phidht$ (Figure~\ref{fig:dc-proxies}(e)) agrees
with the other proxies we consider, but the potential shape along $x$ shows
pronounced $\di$-scale fluctuation that does not appear in the other proxies;
e.g., in the post-shock region $x-x_\mt{sh} = -10$ to $0\di$
and the foreshock region $x-x_\mt{sh}=15$ to $30\di$.
Why is this?
If the off-diagonal terms in $\vec{P}_\mt{e}$ were the \emph{sole} cause for
disagreement between $\phi_\prll$ and $\phidht$, as discussed in
Section~\ref{sec:phiprllmethdht}, then we would expect
Equation~\ref{eq:phiambx} (blue dotted) to coincide with $\phidht$ (orange).
But that is not so.
The disagreement between $\phi_{\prll,\mt{amb}}$ and $\phidht$ must come
from other effects, e.g., $\vec{E}_\mt{HT} \ne \vec{E}_\mt{amb}$ due to
low-frequency motional fluctuations outside the shock ramp that are not fully
cancelled by the global HT frame boost.

Our calculation of Equation~\eqref{eq:phiambx} also shows that the non-diagonal
terms in $\vec{P}_\mt{e}$ contribute appreciably to $\phi_{\prll,\mt{amb}}$ in
our shocks.
The non-diagonal $\vec{P}_\mt{e}$ gradients are responsible for (i) potential
spikes within the shock precursor region $x-x_\mt{sh} = 0$ to $30 \di$, and
(ii) a $\abt 20\%$ increase in $\phi_\prll$ across the shock ramp at
$x=x_\mt{sh}$, compared to integrating only the
$\uvec{x}\cdot(\del\cdot\vec{P}_\mt{e})$ piece
(Figure~\ref{fig:dc-proxies}(c)).
We expect that our shocks, having relatively low $\bp$,
will show larger pressure anisotropies than higher-$\bp$ shocks wherein
pressure anisotropy may be bounded to lower magnitude by various
microinstabilities.

Figure~\ref{fig:dc-proxies-all} proceeds to a wider shock parameter range.
We show the same $\phi_\prll$ proxies for $\thetaBn=75^\circ$ to $55^\circ$ and
varying $\Ms$, following the same procedure as in Figure~\ref{fig:dc-proxies}.
All runs use 50 evenly-spaced snapshots within a short time interval.
For $\Ms=3$--$5$, $t\approx40.02$--$40.12{\Omci}^{-1}$ (with $\pm
0.01{\Omci}^{-1}$ variation on the exact timing for individual runs).
For $\Ms=7$, $t=30.04$--$30.14{\Omci}^{-1}$.
For $\Ms=10$, $t=20.23$--$20.33{\Omci}^{-1}$.
The potential is pinned to $\phi=0$ at $x-x_\mt{sh}=45 \di$ for all shocks and
time snapshots.

What do we learn from Figure~\ref{fig:dc-proxies-all}?
The HT-frame potential is noisy, but it does a reasonable job of replicating
the magnitude of the $\phi_\prll$ jump across the shock ramp, as measured by
other proxies.
The HT-frame potential deviates from other $\phi_\prll$ estimators within
the precursor wave trains of low $\Ms$ and low $\thetaBn$ shocks.
For example, in the $\Ms=3,\thetaBn=55^\circ$ case, the HT potential shows a
gradual rise with less evident fluctuations than $\phi_\prll$.
In the $\Ms=4,\thetaBn=55^\circ$ case, the HT potential shows more fluctuation
than $\phi_\prll$.
Disagreement between $\phidht$ and $\phi_\prll$ is reasonable in such
precursors because the
HT-frame boost's cancellation of
motional electric fields may be imperfect within a shock transition of finite
width.

The particle-based $\phi_\prll$ proxies agree well with each other for
$\thetaBn=65$--$55^\circ$ (Figure~\ref{fig:dc-proxies-all}); more disagreement
is seen in $\thetaBn=75^\circ$ shocks, which are closer to perpendicular and
which we expect to deviate from the quasi-static electron heating model
description \citep{goodrich1984}.
The direct grid measure $\phi_{\prll,\mt{grid}}$ can be noisy for weak shocks,
but our averaging procedure reduces the uncertainty across most of the shock
parameter range considered.
We suggest that agreement or disagreement between our proxy measurements in
Figures~\ref{fig:dc-proxies} and \ref{fig:dc-proxies-all} can be broadly
attributed to systematic error: 2D effects, neglected terms in the generalized
Ohm's law (i.e., effective frictional force from collisionless scattering), or
finite-Larmor radius drift contributions to electron heating that involve
$\vec{B}$-perpendicular electric fields \citep{northrop1963}.

The calculation of Equation~\eqref{eq:phiambx}, which tests whether
$\vec{E}_\mt{amb}$ may be approximated as lying along $\uvec{x}$, returns a
cross-shock potential jump that is lower than the other $\phiprll$ proxies
in all but the weakest ($\Ms=3$) shocks.

Lastly, the Liouville-mapping inferred values for
$\Delta\tilde{\phi}_\prll$ (red triangles) show multiple trends across the
$(\Ms,\thetaBn)$ parameter range.
The reader may inspect the detailed Liouville mapping fits of
$f(\vec{r},\vec{p})$ in Appendix~\ref{app:liouville}.
Weaker, lower-$\thetaBn$ shocks (upper right of
Figure~\ref{fig:dc-proxies-all}) return a Liouville-mapping
$\Delta\tilde{\phi}_\parallel$ value for
incoming electrons that is in general agreement with
Equations~\eqref{eq:phigrid}--\eqref{eq:phidht} and has a range of tens of
percent, when fitted to different bands in $f$ (see
Figure~\ref{fig:liouville-fit-procedure}).
The inferred potential for outgoing ($p_\prll>0$) electrons is consistently
smaller than that for incoming ($p_\prll<0$) electrons.
Towards stronger and higher-$\thetaBn$ shocks, the asymmetry between
outgoing- and incoming-inferred potentials becomes larger, and the
Liouville-inferred potential jump may disagree with
Equations~\eqref{eq:phigrid}--\eqref{eq:phidht} by a factor of $2\times$ or
more.
In cases where the outgoing ($p_\prll >0$) electrons return a value of
$\Delta\phiprll$ significantly smaller than other estimators such as
$\phiamb$, the Liouville mapping may not have a straightforward physical
interpretation.
Downstream electrons with $p_\prll > 0$ could be well confined by $\phiprll$,
and other mechanisms in the shock ramp may be needed to explain the
outgoing upstream electron distributions.
In the $\Ms=10$, $\thetaBn=75^\circ$ case (Figure~\ref{fig:dc-proxies-all},
bottom left), no Liouville mapping is performed for outgoing electrons
because too few upstream electrons have HT-frame $p_\prll > 0$.

In Figure~\ref{fig:dc-proxies-all}, we have omitted the near-luminal
$\thetaBn=85^\circ$ shocks.
In these shocks, it is less meaningful to describe electron heating in terms of
a global, 1D parallel potential.
Conduction along $\uvec{x}$ becomes comparable to, or slower than, bulk advection of
flux tubes into the shock.
And, we have not accounted for finite-Larmor radius drifts in the electron
heating model \citep{goodrich1984}.
Electrons will still gain energy by parallel electric fields in sufficiently
strong shocks, but this parallel energization becomes a local process that is
not well described by a 1D, $y$-averaged potential \citep{tran2020}.

Based on Figures~\ref{fig:dc-proxies} and \ref{fig:dc-proxies-all}, we adopt
$\phiamb$ as our preferred estimator for $\phi_\parallel$ in the rest of this
manuscript, as it can provide a reasonable estimate of $\phiprll$ from a single
time snapshot.

\section{Cross-shock Potential Scaling with Shock Parameters}
\label{sec:phiprll-scaling}

\subsection{Parallel Potential Scaling}

\begin{figure}
    \includegraphics[width=3.375in]{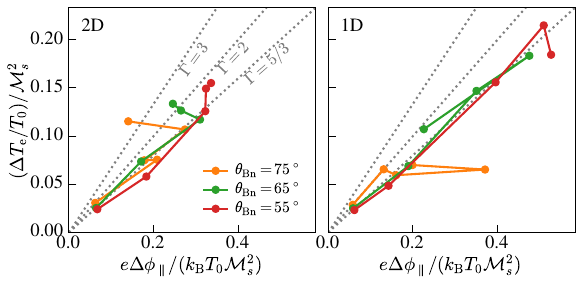}
    \caption{
        Electron temperature jump $\Delta \Te$ versus parallel potential jump
        $\Delta \phiprll$, in units of $T_0 \Ms^2$.
        Each marker represents a single shock simulation.
        Orange, green, red markers represent
        $\thetaBn = 75$, $65$, $55^\circ$ respectively; markers within each
        color set correspond to varying $\Ms$, with low $\Ms$ at the
        bottom-left corner of each plot.
        Gray dotted lines correspond to a fluid approximation,
        $\Delta T_\mt{e} = \Delta\phi_\prll (\Gamma-1) /\Gamma$,
        with varying $\Gamma=3$, $2$, and $5/3$.
        Left panel shows 2D shocks; right panel shows 1D shocks.
    }
    \label{fig:deltapot-scatter-te}
\end{figure}

\begin{figure}
    \includegraphics[width=3.375in]{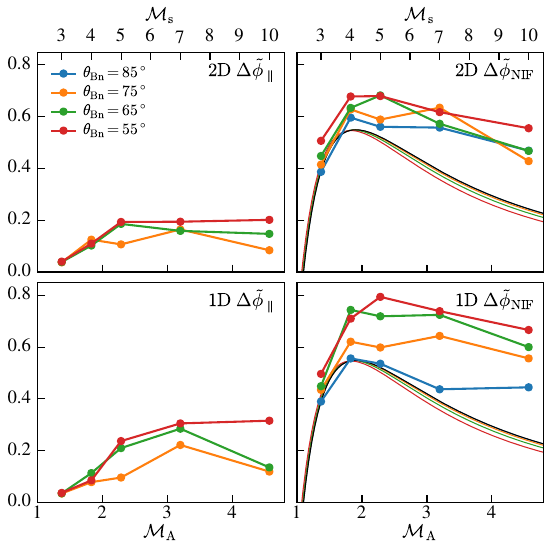}
    \caption{
        Parallel potential jump $\Delta\tilde{\phi}_\prll$ (left column) and
        NIF potential jump $\Delta\tilde{\phi}_\mt{NIF}$ (right column) as a
        function of Mach number.
        Top and bottom rows present 2D and 1D shocks respectively.
        The marker scheme is identical to Figure~\ref{fig:deltapot-scatter-te},
        except that we add $\thetaBn=85^\circ$ shock data to the
        $\Delta\tilde{\phi}_\mt{NIF}$ plot.
        The NIF potential jumps are compared to
        the low-$\Ma$ model Equation~\eqref{eq:gedalin} (solid curved
        lines) with varying $\thetaBn=85$--$55^\circ$; colors match the
        marker sets.
    }
    \label{fig:deltapot-scatter-mach}
\end{figure}

What is the relationship between $\phi_\prll$ and $T_\mt{e}$ as a
function of shock parameters?
If we assume isotropic electrons with a polytropic equation of state $P_\mt{e}
\propto n_\mt{e}^\Gamma$, where $\Gamma$ is an effective adiabatic index,
Equations~\eqref{eq:phiamb} and \eqref{eq:phigyr} simplify to
\[
    \Delta \phiprll = \frac{\Gamma}{\Gamma-1} \kB \Delta T_\mt{e} \, .
\]
The effective adiabatic index $\Gamma$ should reflect the underlying kinetic
physics and simulation setup (e.g., a local planar shock versus a global
magnetosphere model), so what is its value?

The correlation between the cross-shock jumps $\Delta\phi_\prll$ and
$\Delta \Te$ has been previously estimated using ISEE data at Earth's bow shock
\citep{schwartz1988,hull2000-isee}, MAVEN data at Mars
\citep{horaites2021,xu2021}, and various hybrid and PIC simulations
\citep{thomsen1987-noncoplanar,savoini1994,nishimura2002}.
\citet{schwartz1988} report a sample average of $\Gamma=2.7\pm1.9$ and further
suggest a progression from $\Gamma=5/3$ for weak, subcritical shocks up to
$\Gamma=3$ for strong, supercritical shocks, where sub/super-critical follows
the \citet{marshall1955,edmiston1984} criterion.
\citet{hull2000-isee} report $\Gamma=2\pm0.1$ from a linear fit between
$\Delta \Te$ and $\Delta \phidht$.
\citet{savoini1994} estimated $\Gamma=2.9$ for a 2D PIC shock with
$\thetaBn=55^\circ$.
The correlation between $\Delta\phi_\prll$ and $\Delta \Te$ was studied by
\citet{nishimura2002} for 1D quasi-parallel PIC shocks, but it has not been
assessed, to our knowledge, for a suite of 2D
quasi-perpendicular
PIC simulations.

We measure the cross-shock potential and electron temperature jumps in the
post-shock region $x-x_\mt{sh}=-18$ to $-3\di$ as space- and density-weighted
averages respectively (Appendix~\ref{app:deltapot-measure}).
Besides $\phiprll$, we also measure the normal incidence frame (NIF)
potential $\phi_\mt{NIF} = \int E_x \dtl x$ as a commensal ``add-on'';
$\phi_\mt{NIF}$ is not directly related to our discussion of electron
heating, but it is convenient to also present here.

Figure~\ref{fig:deltapot-scatter-te} suggests that both 1D and 2D
simulations show an effective $\Gamma \approx 5/3$ relation between
$\Delta\phiprll$ and $\Delta T_\mt{e}$, although the stronger shocks may
deviate towards a higher effective $\Gamma$.
The 1D simulations can attain higher $\Te$ and $\Delta\phiprll$, yet the
effective $\Gamma=5/3$ remains similar to our 2D simulations.
We caution that the effective $\Gamma$ for post-shock electrons could be
sensitive to our choice of domain left-side boundary condition, which
may modify the post-shock electron isotropization.

The left column of Figure~\ref{fig:deltapot-scatter-mach} shows that in
2D shocks, $\Delta\tilde{\phi}_\prll$ increases with Mach number until
$\Ma \sim 2$--$3$, after which $\Delta\tilde{\phi}_\prll$ saturates around
$0.1$--$0.2$.
For 1D shocks, $\Delta\tilde{\phi}_\prll$ can achieve somewhat larger values,
up to $\abt0.3$ at $\thetaBn = 55^\circ$.
The normalized value of $\Delta\tilde{\phi}_\prll\sim 0.1$--$0.2$ is
consistent with spacecraft data \citep{schwartz1988,hull2000-isee,xu2021}, as
well as prior simulations \citep{thomsen1987-noncoplanar,savoini1994} and
theoretical expectations \citep{goodrich1984,thomsen1987-noncoplanar}.

\subsection{Normal Incidence Frame Potential Scaling}

We further consider how the normal incidence frame (NIF) cross-shock potential,
$\phi_\mt{NIF} = \int E_x \dtl x$, scales with $\Mms$ and $\thetaBn$ in our
simulations.
Although less relevant to electron heating, $\phi_\mt{NIF}$ is still
important to a kinetic shock's internal structure \citep{burgess2015}.
The NIF cross-shock potential's dependence on Mach number has also been studied
in prior hybrid \citep{leroy1982,quest1986} and PIC \citep{shimada2005}
shock simulations.

Our results are shown in the right column of
Figure~\ref{fig:deltapot-scatter-mach}, which compares $\Delta\phi_\mt{NIF}$ to
a low-$\Ma$ model from \citet[Equation~(40)]{gedalin1996} (and
\citet[Equation~(28)]{gedalin1997}; \citet[Sec.~5.1]{jones1991}),
\begin{equation} \label{eq:gedalin}
    \phi_\mt{NIF} = \frac{2(r-1)(1+\beta_\mt{e})}{\Ma^2} \, .
\end{equation}
Equation~\eqref{eq:gedalin} comes from integrating $E_x$ of generalized
Ohm's Law across the shock, assuming
scalar $P_\mt{e} \propto n_\mt{e}^2$ and $B \propto n_\mt{e}$.
The upstream $\beta_\mt{e} = \bp/2$ in our simulations.
To evaluate the compression ratio $r$ in Equation~\eqref{eq:gedalin}, we
use the oblique MHD shock jump conditions, similar to \citet{bale2008}.
Our calculation of $r$ introduces a weak $\thetaBn$ dependence
shown by the closely-spaced curves of varying color in
Figure~\ref{fig:deltapot-scatter-mach}; an oblique $\thetaBn=55^\circ$
decreases the model prediction by $\lesssim 10\%$ compared to the
nearly-perpendicular $\thetaBn=85^\circ$.

Figure~\ref{fig:deltapot-scatter-mach} shows that $\Delta\phi_\mt{NIF}$ is of order
one-half the incident ion bulk kinetic energy, $0.5 \mi u_\mt{sh}^2$, as is well
understood \citep{leroy1982,leroy1983,burgess2015}.
Both 2D and 1D shocks show that $\Delta\phi_\mt{NIF}$ varies with $\thetaBn$, with
more oblique $\thetaBn \sim 55^\circ$ yielding higher $\Delta\phi_\mt{NIF}$
than more perpendicular $\thetaBn$.
The 1D shocks also show stronger variation in $\Delta\phi_\mt{NIF}$ with
$\thetaBn$.
We observe that going from $\Ms=3$ to $4$, the potential increases for all
$\thetaBn$ considered.
Towards higher $\Ms$, the potential
appears to level off or decrease slightly; the \citet{gedalin1997} model does
not apply for these stronger shocks.

\begin{figure*}
    \includegraphics[width=\textwidth]{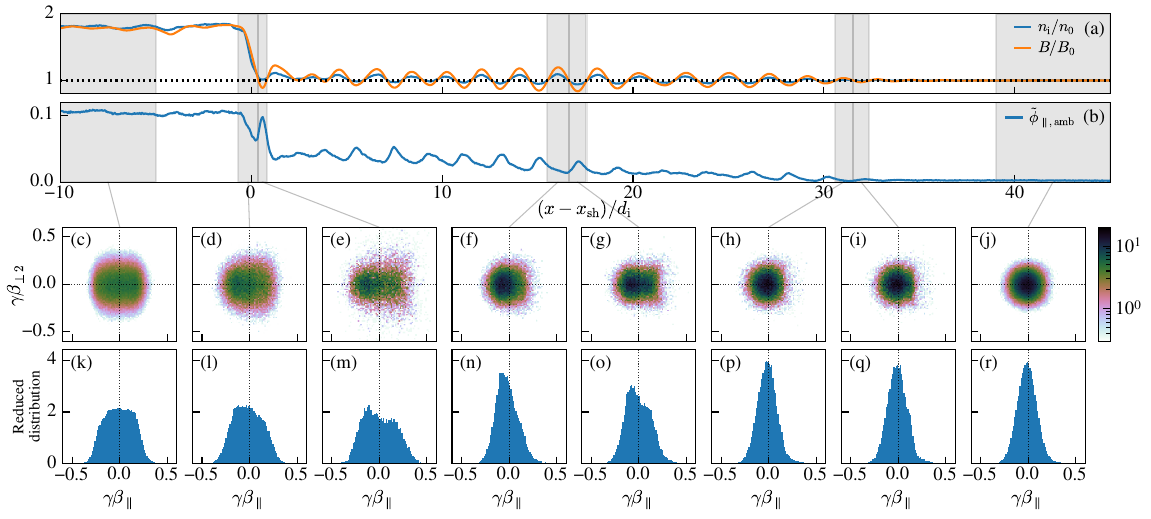}
    \caption{
        Electron momentum distribution functions sampled in intervals along $x$
        for a 2D, $\Ms=4$, $\thetaBn=65^\circ$ shock at $t=40\Omci^{-1}$.
        (a): Ion density $n_\mt{i}$ and total magnetic field strength $B$
        normalized to their upstream values, $n_0$ and $B_0$.
        (b): Parallel potential $\tilde{\phi}_{\prll,\mt{amb}}$ along $x$
        computed as in Equation~\eqref{eq:phiamb} and normalized to
        $0.5 \mi u_\mt{sh}^2$.
        (c)--(j): Reduced 2D distributions in parallel and perpendicular
        momenta, $\gamma\beta_\prll$ and $\gamma\beta_{\perp 2}$, normalized so
        that the histogram integral is unity.
        The normalization thus does not show variation in electron density
        $n_\mt{e}$ along the shock.
        Momentum coordinate system is defined in manuscript text.
        (k)--(r): Reduced 1D distribution in parallel momentum
        $\gamma\beta_\prll$.  Like in panels (c)--(j), the 1D histogram bins
        are normalized so that their integral equals one.
        The electron distributions of (c)--(r) are sampled from gray-shaded $x$
        intervals in (a)--(b), shown by light gray lines connecting panels
        (c)--(j) to regions in (b).
    }
    \label{fig:vdf-salami}
\end{figure*}

\begin{figure*}
    \includegraphics[width=\textwidth]{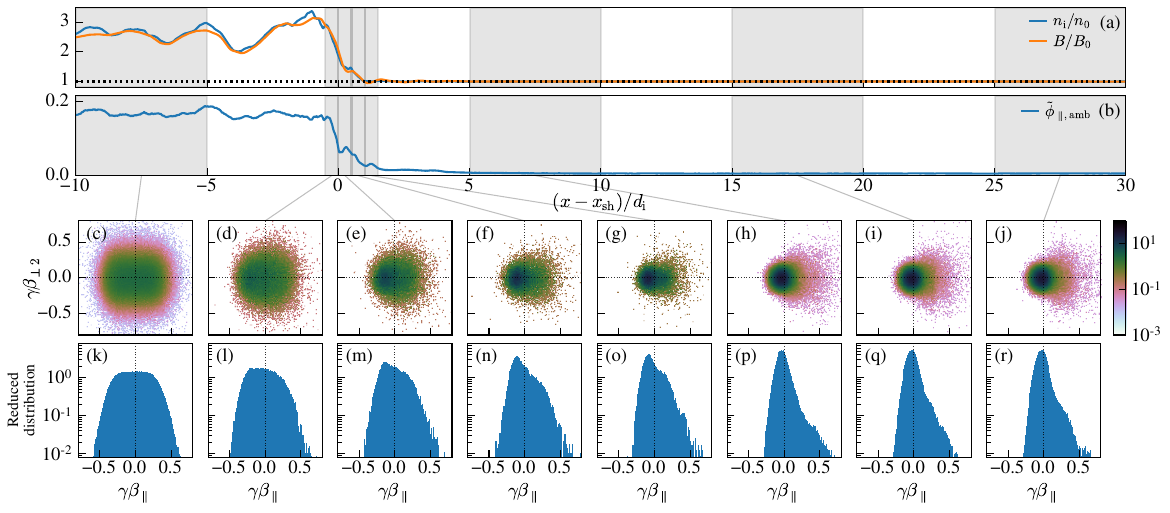}
    \caption{
        Electron momentum distribution functions sampled in intervals along $x$
        for a 2D, $\Ms=7$, $\thetaBn=65^\circ$ shock at $t=30\Omci^{-1}$.
        See Figure~\ref{fig:vdf-salami} for panel explanation.
    }
    \label{fig:vdf-salami-Ms7}
\end{figure*}

\begin{figure*}
    \includegraphics[width=\textwidth]{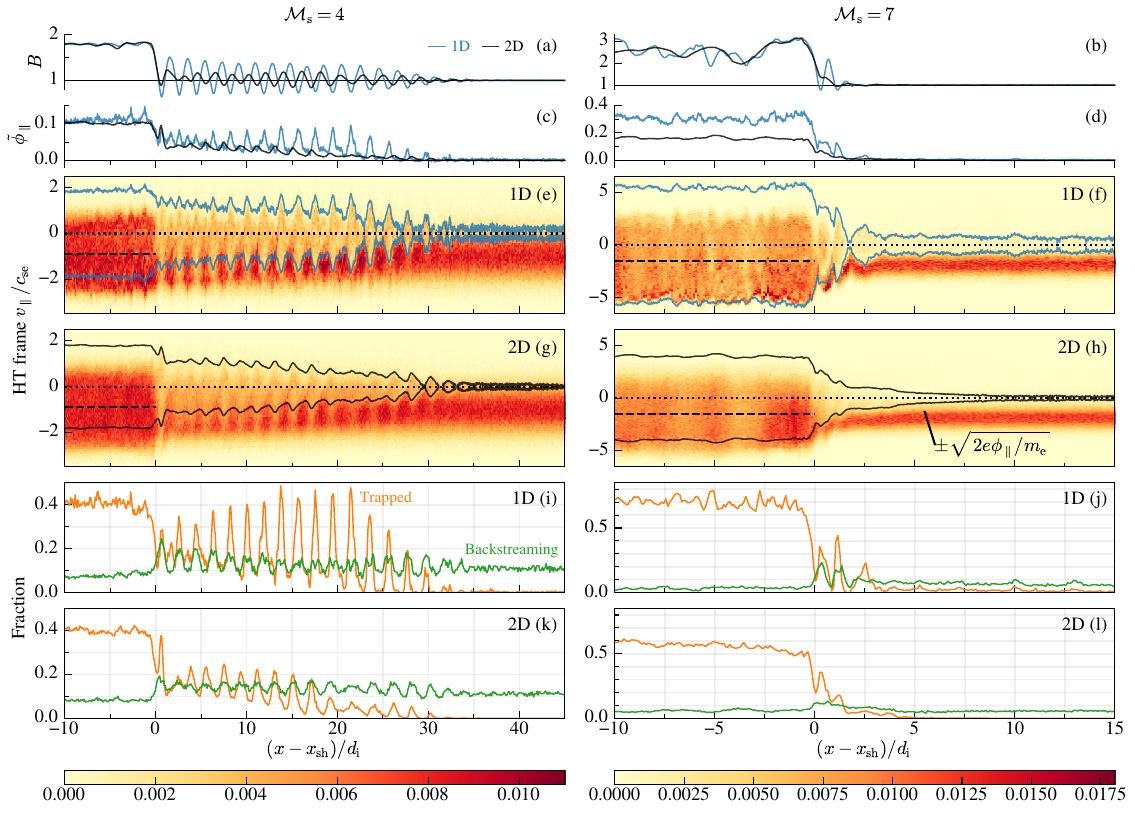}
    \caption{
        Electron $x$-$v_\prll$ phase space behavior in the HT frame for
        $\Ms=4$ (left column) and $\Ms=7$ (right column) shocks with
        $\thetaBn=65^\circ$, both 1D and 2D cases.
        (a--b): Magnetic field strength $B/B_0$.
        (c--d): Parallel potential $\tilde{\phi}_{\prll} = \tilde{\phi}_{\prll,\mt{amb}}$.
        (e--h): Electron parallel velocity distributions, normalized to
        upstream thermal speed $c_\mt{se} = \sqrt{\Gamma \kB T_0/\me}$, for the
        1D (e--f) and 2D (g--h) shocks.
        Solid blue (e--f) and black (g--h) curves show the effective potential
        floor or barrier $\pm\sqrt{2 e\phi_\prll/m_\mt{e}}$.
        Black dotted line marks $v_\prll=0$.
        Black dashed line marks HT bulk post-shock velocity $u_\mt{HT2}$
        (Equation~\eqref{eq:udht2}).
        (i--l): Trapped (orange) and backstreaming (green) electron fractions
        for the 1D (i--j) and 2D (k--l) shocks, see text for definitions.
    }
    \label{fig:xvprll}
\end{figure*}

\section{Case Studies of Electron Dynamics Within the Parallel Potential}
\label{sec:case-study}

\subsection{Electron Phase Space}

What electron distributions form within and generate the parallel potential?
We focus upon two shock cases that are weak ($\Ma<\Mw$, sub-critical) or strong
($\Ma>\Mw$, super-critical) and therefore have quite different
structure.
The weaker case is the $\Ms=4$, $\thetaBn=65^\circ$ shock previously shown in
Figures~\ref{fig:v3iy} and \ref{fig:dc-proxies}.
The stronger case is the $\Ms=7$, $\thetaBn=65^\circ$ shock.
Within Section~\ref{sec:case-study}, we refer to each case by its sonic Mach
number $\Ms$ alone.

Figure~\ref{fig:vdf-salami} surveys electron phase space evolution along the
shock-normal coordinate for our 2D $\Ms=4$ case, in the manner of
\citet{scudder1986-i} and \citet{chen2018}.
We show electron momenta $\gamma\beta$ in units of $\me c$, where $\gamma$ is
the relativistic Lorentz factor and $\beta$ is three-velocity in units of $c$.
The momentum coordinates $(\uvec{n}_\prll,\uvec{n}_{\perp 1},\uvec{n}_{\perp 2})$
are an orthogonal, right-handed triad defined by $\uvec{n}_\prll = \uvec{b}$,
$\uvec{n}_{\perp 1} \propto (\uvec{b}\times\vec{V}_\mt{e})\times\uvec{b}$,
and $\uvec{n}_{\perp 2} \propto \uvec{b}\times\vec{V}_\mt{e}$;
here, $\vec{V}_\mt{e}$ is local electron bulk three-velocity, and $\uvec{b}$ is
computed locally at each particle position.
This momentum coordinate system follows \citet{chen2018}.
By construction, $\gamma\beta_{\perp 1}$ includes both bulk motion and Larmor
gyration, and $\gamma\beta_{\perp 2}$ includes only Larmor gyration.

Figure~\ref{fig:vdf-salami}(a) shows that ion density oscillates in sync with
magnetic field strength, as expected for the compressible, fast
(whistler-branch) mode.
In panel (b), we adopt $\phiamb$ as our proxy for the parallel potential.
Panels (c)--(r) show how electron momentum distributions evolve through the
shock, with momenta measured in the simulation frame.
The distributions are thermal Maxwellians far upstream of the shock,
at $x-x_\mt{sh} = 40$--$45\di$ (panels (j),(r)).
Within the shock precursor, the distributions become asymmetric in
$\gamma\beta_\prll$; we see a beam-like component with $\gamma\beta_\prll < 0$
on top of a broader, ``flat-top'' component; the asymmetric form is most
prominent in strong magnetic troughs
with $\delta B/B_0 \abt 0.1$
(panels (m),(o)).
Distributions within magnetic peaks
are closer to a single component in form, but with a skew towards negative
$\gamma\beta_\prll$ (panels (n),(p)).
Downstream of the shock, the beam-like component disappears and the
distribution shows a smooth, flat region in $\gamma\beta_\prll$ centered near
zero (panels (c),(k)).

In some 2D momentum distributions (panels (e),(g)), particles with
$\gamma\beta_\prll > 0$ show a hint of anisotropy, with some particles at
$\gamma\beta_\prll \sim 0.25$ having large $\gamma\beta_{\perp 2} \gtrsim
\gamma\beta_\prll$; the right side of the 2D distributions is stretched along
the $\gamma\beta_{\perp 2}$ axis.
We expect that these back-streaming electrons were magnetic mirror-reflected
within the shock ramp or precursor.
Back-streaming electrons are also visible in the far-upstream 1D distributions
(panels (p),(q),(r)) when plotted with a logarithmic scale.

The evolution of parallel distributions through the shock, showing a transient
beam and flat-top structure, is broadly consistent with satellite observations
\citep{feldman1983-earth,feldman1983-interplanet} and prior 1D and 2D PIC
simulations \citep{savoini1994}.
It can be noted that \citet{feldman1983-interplanet} showed that weak,
interplanetary shocks with $\Ms\sim 1$--$4$ exhibit less beam/flat-top
structure as compared to the stronger bow shock at Earth's magnetosphere.
Although our $\Ms=4$ case has $\Ma<\Mw$ and is sub-critical, it shows parallel
electron behavior similar to
space measurements of
stronger shocks.
This is partly an artifact of the reduced mass ratio adopted for our
simulations; we return to this point in Section~\ref{sec:mime-dgamr}.

Electron distributions in the $\Ms=7$ case (Figure~\ref{fig:vdf-salami-Ms7})
show similar distortions as the $\Ms=4$ case:
namely, a one-sided beam with $\gamma\beta_\prll < 0$ accelerating towards the
shock and eroding through the shock ramp to leave a broad, flattened post-shock
distribution in $\gamma\beta_\prll$.
The population of
back-streaming
electrons is more prominent than in the $\Ms=4$ case.

Figure~\ref{fig:xvprll} shows how the HT-frame $\gamma\beta_\prll$ electron
distributions correlate with the potential $\phi_\prll$ for the $\Ms=4$ and
$\Ms=7$ case studies in 1D and 2D.
In Figure~\ref{fig:xvprll}(e--h), the blue and black solid curves show $\phiprll$ recast as a
parallel velocity $\pm \sqrt{2 e\phi_\prll/\me}$, and black dashed lines show
the approximate HT-frame downstream bulk velocity,
\begin{equation} \label{eq:udht2}
    u_\mt{HT2} = \frac{u_\mt{sh}}{r} \left( 1 + (r^2-(u_\mt{sh}/c)^2) \tan^2 \thetaBn \right)^{1/2}
    \, .
\end{equation}
In Figure~\ref{fig:xvprll}(i--l), trapped electrons are defined to have total
energy $\varepsilon < 0$ (Equation~\eqref{eq:eom-ener}), while backstreaming
electrons have $\varepsilon > 0$ and $v_\prll > 0$.
The electron fractions in Figure~\ref{fig:xvprll}(i--l) are computed within
$0.1\di$ bins along $x$ using the 1D $y$-averaged $\phiprll(x)=\phiamb(x)$.
Some downstream electrons with $v_\prll < 0$ are not
counted as trapped in Figure~\ref{fig:xvprll}, but most such electrons are
\emph{de facto} trapped because reflection at the left boundary or scattering
within the downstream rest frame generally will not give them $\varepsilon > 0$
and $v_\prll > 0$ as needed to escape back upstream.

We may understand an electron's parallel velocity evolution by casting
Equations~\eqref{eq:eom-ener} and \eqref{eq:eom-mu} into dimensionless form:
\begin{equation}
    \label{eq:vprll-evol}
    \frac{v_\prll^2 - v_{\prll 0}^2}{c_{se}^2}
    =
    \Ms^2
    \tilde{\phi}_{\prll}(x)
    - \frac{v_{\perp 0}^2}{c_{se}^2} \left(\frac{B(x)}{B_0} - 1\right)
\end{equation}
with $v_{\prll 0}$ and $v_{\perp 0}$ being the electron's initial HT-frame
velocity components far upstream, where $\phiprll(x)=0$ and $B(x)=B_0$.
The upstream electron thermal speed $c_{se} = \sqrt{\Gamma \kB T_0/\me}$, and
the normalized $\tilde{\phi}_\prll = e\phiprll / (0.5 \mi u_\mt{sh}^2)$ is
typically $0.1$--$0.2$ in our simulated shocks
(Figure~\ref{fig:deltapot-scatter-mach}).
Equation~\eqref{eq:vprll-evol} shows that electrons gain parallel energy from
$\phi_\prll$ and lose energy from magnetic mirroring, and $\Ms^2$ moderates the
relative importance of $\phiprll$ versus mirroring.

Let us trace how electrons evolve within the potential $\phi_\prll$ and the
corresponding velocity floor/barrier $\pm\sqrt{2 e\phi_\prll/\me}$, neglecting
scattering.
Suppose an electron enters the shock with $v_{\prll 0} < 0$ and
$v_{\perp 0} = 0$ (no magnetic mirroring) in the HT frame; further suppose that
$|v_{\prll 0}| \ll \sqrt{2e\phi_\prll/\me}$.
That electron will follow the negative potential floor
$v_\prll(x) = - \sqrt{2 e\phi_\parallel(x)/\me}$ into the shock.
It reflects at the left boundary by reversing the sign of $v_x$ in the
simulation frame, i.e. the downstream rest frame.
The post-shock $v_\prll$ distribution is thus centered on the HT bulk flow
velocity $v_\prll = -u_\mt{HT2}$ (Equation~\eqref{eq:udht2}), shown by dashed
horizontal lines in Figure~\ref{fig:xvprll}(e--h).
Post-shock electrons may escape upstream if they have
$v_\prll \gtrsim \sqrt{2 e\phi_\prll/\me}$ (more precisely, $\varepsilon > 0$
for non-zero pitch angle).
The fraction of downstream electrons that may escape upstream is $\lesssim
10\%$ of the total downstream population in all case-study shocks
(Figure~\ref{fig:xvprll}(i--l)); most shocked electrons are confined to
$x-x_\mt{sh} < 0$, consistent with the prediction of Figure~\ref{fig:regimes}.

If an electron entering the shock instead has non-zero $v_\perp$ and a
large-enough pitch angle, it may mirror reflect specularly within the HT frame
(i.e., $v_\prll\to-v_\prll$).
If the macroscopic $B(x)$ and $\phi_\prll(x)$ are time-stationary and there is
no scattering, the reflected electron will freely escape back upstream with
$\varepsilon > 0$ and $v_\prll > 0$.
In practice, reflection combined with scattering may lead to local electron
trapping and energization within the precursor wave train and shock ramp
\citep{katou2019}.

In each panel of Figure~\ref{fig:xvprll}, the potential floor (solid blue and
black lines with $v_\prll < 0$) tracks the deformation of the upstream thermal
electron beam entering the shock from right to left.
In the post-shock region, the potential floor roughly corresponds to the
anti-parallel ($v_\prll<0$) edge of the electron distribution.
The agreement of the potential floor and the electron distribution edge in
Figure~\ref{fig:xvprll}
corresponds to good agreement at $p_\prll < 0$ in the Liouville mapping
procedure of Figure~\ref{fig:liouville-fit-procedure}.

Let us now compare 1D versus 2D for the $\Ms=4$ case in Figure~\ref{fig:xvprll}.
The overall structure of $\phi_\prll$ is similar in both 1D and 2D, with a
gradual rise in the shock precursor region and jump at the shock ramp to a
final post-shock value of $\Delta\tilde{\phi}_\prll \approx 0.1$.
The 1D shock shows sharper $\di$-scale potential spikes embedded within the
precursor wave train; in both 1D and 2D the spikes occur within magnetic
troughs.
Electrons clump at local magnetic maxima, especially in 1D, which we
interpret as due to magnetic mirroring.
In 1D, $\phi_\prll$ spikes also occur at magnetic maxima within $x-x_\mt{sh} =
0$--$5\di$ (Figure~\ref{fig:xvprll}, top left panel) just ahead of the shock
ramp.

The 1D and 2D cases differ in the post-shock region $x < x_\mt{sh}$ for both
the $\Ms=4$ and $7$ cases.
The 2D-shock distributions are diffuse and smooth in $v_\prll$, whereas the
1D-shock distributions show localized (in $x$) beams or clumps
that may correspond to transient phase space holes.
The 1D case also shows larger precursor fluctuations in the trapped and
untrapped electron fractions (and, larger magnetic and $\phi_\prll$
fluctuations) than the 2D case.
It is possible that similarly strong fluctuations occur in 2D but are hidden by
$y$-averaging; however, inspection of $B/B_0$ images suggests that the 2D
precursor fluctuations are coherent and so of lower amplitude than in 1D.

In the $\Ms=7$ case, the 1D jump in $\phi_\prll$ is $\abt 2\times$
larger than the equivalent 2D shock.
The 2D shock shows more $v_\prll$ diffusion in the shock foot as electrons
approach the ramp ($x-x_\mt{sh} \sim 1\di$).
We suggest that strong, non-adiabatic scattering embedded within the shock ramp
drives so-called ``infilling'' of the parallel distribution \citep{hull1998}
and hence a net cooling, as compared to the 1D case.

\begin{figure}
    \includegraphics[width=3.375in]{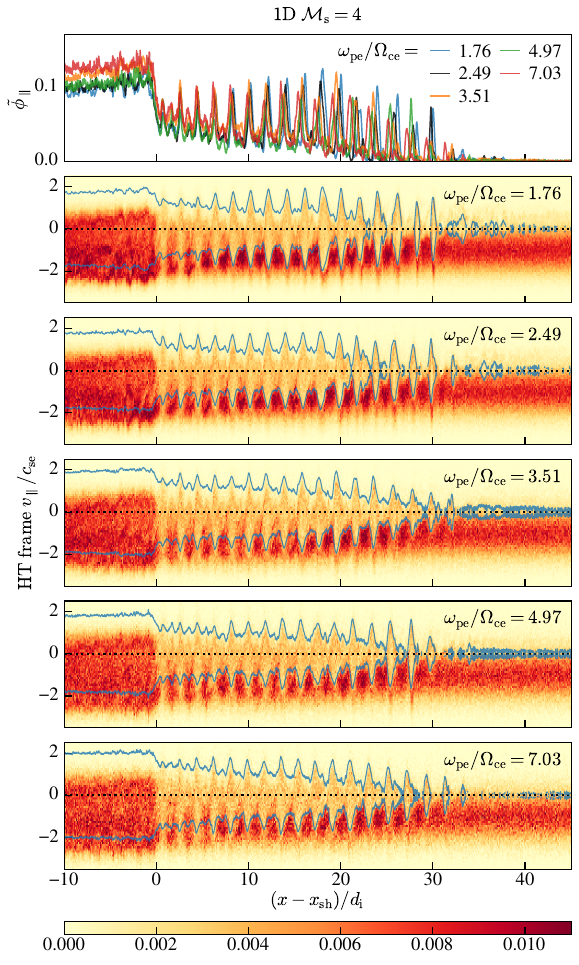}
    \caption{
        Effect of $\ompe/\Omce$, ranging from 1.76 to 7.03, upon $\phiprll(x)$
        and electron $x$-$v_\prll$ phase space behavior in the HT frame for 1D
        $\Ms=4$, $\thetaBn=65^\circ$ shock.
        Top panel: $\tilde{\phi}_\prll$ with colors indicating $\ompe/\Omce$
        value.
        Fiducial $\ompe/\Omce=2.49$ is black curve.
        Bottom panels: electron $x$-$v_\prll$ phase space, same organization as
        Figure~\ref{fig:xvprll}(e), with $\ompe/\Omce$ increasing as rows
        descend.  Third row from top is fiducial case, like
        Figure~\ref{fig:xvprll}(e)
        but simulated using $4\times$ more particles.
    }
    \label{fig:xvprll-dgamr}
\end{figure}

\begin{figure}
    \includegraphics[width=3.375in]{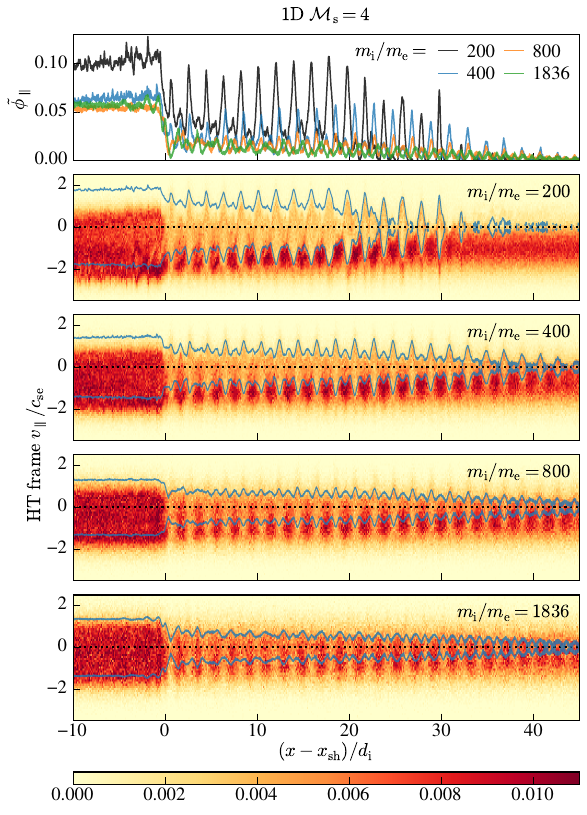}
    \caption{
        Effect of $\mime$, ranging from 200 to 1836, upon $\phiprll(x)$ and
        electron $x$-$v_\prll$ phase space behavior in the HT frame for 1D
        $\Ms=4$, $\thetaBn=65^\circ$ shock.
        The upstream electron temperature $\delgame=0.01$ for all runs.
        Top panel: $\tilde{\phi}_\prll$ with colors indicating $\mime$ value.
        Fiducial $\mime=200$ is black curve.
        Bottom panels: electron $x$-$v_\prll$ phase space, same organization as
        Figure~\ref{fig:xvprll}(e), with $\mime$ increasing as rows
        descend.  Second row from top is fiducial case, like
        Figure~\ref{fig:xvprll}(e)
        but simulated using $4\times$ more particles.
    }
    \label{fig:xvprll-mime}
\end{figure}

\subsection{Effects of Electron Plasma-Cyclotron Frequency Ratio and Ion-Electron Mass Ratio}
\label{sec:mime-dgamr}

Let us now explore the effects of the electron plasma-to-cyclotron frequency
ratio $\ompe/\Omce$ and the mass ratio $\mime$ upon the parallel potential, for
fixed Mach number and $\bp$.
In this Subsection, we focus solely on the 1D $\Ms=4$ case for two reasons.
First, the extended $\phiprll$ structure over many $\di$ in the $\Ms=4$ shock
precursor has not been studied before in a quasi-perpendicular shock, to our
knowledge.
Second, computing cost
and numerical noise both
rise as either $\mime$ or $\ompe/\Omce$ are raised towards realistic values;
the use of 1D simulations helps us limit both cost and noise.
In this Subsection (Figures~\ref{fig:xvprll-dgamr}, \ref{fig:xvprll-mime})
all simulations use $8192$ upstream particles per cell, four times larger
than our standard runs (Section~\ref{sec:shock-param}).

Figure~\ref{fig:xvprll-dgamr} shows that $\ompe/\Omce$ does not have a strong
effect on $\tilde{\phi}_\prll$ or the electron phase-space behavior,
which we interpret as follows.
Suppose that $\phiprll$ scales with some electrostatic fluctuation strength,
$\delta E_\prll$, and is integrated over an ion skin depth:
\[
    e \phi_\prll \sim e \delta E_\prll \di
    \, .
\]
When normalized to ion kinetic energy,
\[
    \tilde{\phi}_\prll =
    \frac{2 e \phi_\prll}{\mi u_\mt{sh}^2}
    \sim \frac{2}{\Ma} \left( \frac{\delta E_\prll}{u_\mt{sh} B_0/c} \right)
    \, .
\]
Let us now suppose that $\delta E_\prll$ scales like a whistler wave's
electromagnetic fluctuation $\delta E_\perp$, which will be motional for low
frequencies: $\delta E_\perp \sim \delta u \delta B/c$ (where $\delta u$ and
$\delta B$ are the wave's velocity and magnetic fluctuations).
Then
\[
    \delta E_\prll
    \sim \delta E_\perp
    \sim \left(\frac{\delta u}{\vA}\right)
         \left(\frac{\delta B}{B_0}\right)
         \frac{1}{\Ma}
         \frac{u_\mt{sh} B_0}{c}
    \, .
\]
The fluctuations $\delta u/\vA$ and $\delta B/B_0$ should not depend on
$u_\mt{sh}/c$ for the non-relativistic solar wind \citep{verscharen2020}.
Then,
\[
    \tilde{\phi}_\prll
    \sim \frac{2}{\Ma^2}
        \left(\frac{\delta u}{\vA}\right)
        \left(\frac{\delta B}{B_0}\right)
    \, .
\]
In this scaling,
we find that $\tilde{\phi}_\prll$ shows no explicit dependence on $u_\mt{sh}/c$
or $\ompe/\Omce$.
Our heuristic argument has caveats.
The whistler wave strength $\delta B/B_0 > 0.1$ is outside the regime of linear
fluctuations, especially at the shock ramp ($\delta B/B_0 \sim 1$).
And, the supposed proportionality $\delta E_\prll \sim \delta E_\perp$ is
suspect; the strength of $\delta E_\prll$ could be regulated by secondary
electrostatic modes that could introduce $\mi/\me$ or $\ompe/\Omce$ dependence
into the scaling argument.

Figure~\ref{fig:xvprll-mime} shows with increasing mass ratio $\mime$:
$\phiprll$ decreases, the electron $v_\prll$ phase space is less distorted, and
magnetic mirroring from $\mu$ conservation becomes more important.
Both downstream and upstream distributions are more centered on $v_\prll = 0$
with increasing $\mime$.
Is the changed electron response due solely to the decrease in $\phiprll$?
Or, would the electron response change even if $\phiprll(x)$ were identical for
both $\mime=200$ and $1836$?

Suppose that we have two shocks with the same $\Ms$, $\tilde{\phi}_\prll(x)$,
and $B(x)$ profiles, and the shocks differ only in their mass ratio $\mime$.
The right-hand side of Equation~\eqref{eq:vprll-evol} is not directly affected
by $\mime$, since $(v_{\perp 0} / c_{se})^2 \sim 1$ in the HT frame, so the
relative importance of $\phiprll$ and magnetic mirroring is unchanged.
In other words, the Liouville mapping of a single trajectory starting at any
$(v_{\prll 0}, v_{\perp 0})$ does not change with the mass ratio.
Instead, $\mime$ influences the electron dynamics via $v_{\prll 0}$.
When the HT-frame bulk upstream velocity $u_\mt{HT} = u_\mt{sh}/\cos\thetaBn$
is much larger than the electron thermal speed $c_{se}$, then
\begin{equation}
    \label{eq:vprll-dht-scaling}
    \frac{v_{\prll 0}}{c_{se}}
    \sim \frac{u_{HT}}{c_{se}} = \frac{\Ms}{\cos\thetaBn} \sqrt{\frac{\me}{\mi}}
    \, .
\end{equation}
In the opposite limit $u_\mt{HT} \ll c_{se}$, $v_{\prll 0}/c_{se} \sim 1$.
The mass ratio $\mime$ shifts the distribution along $v_{\prll 0}/c_{se}$ in
the HT frame.
Although $\mime$ does not change electron trajectories and separatrices in
phase space, which are fully determined by $\tilde{\phi}_\prll(x)$ and $B(x)$
(Equation~\eqref{eq:vprll-evol}), $\mime$ does change the initial sampling of
said trajectories, and hence the relative fraction of electrons that pass into
the shock downstream as opposed to being reflected by magnetic mirroring.

\begin{figure}
    \includegraphics[width=3.375in]{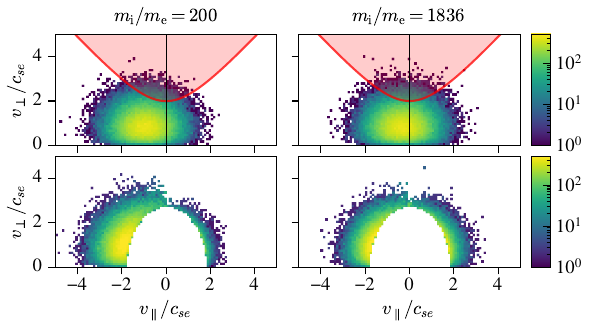}
    \caption{
        Liouville mapping for $B/B_0=1.8$ and $\tilde{\phi}_\prll = 0.10$,
        holding $\Ms=4$, $\thetaBn=65^\circ$ fixed.
        We take $10^5$ samples of a Maxwellian, boost into the HT frame, and
        map electron trajectories from upstream to downstream; all velocities
        and boosts are assumed non-relativistic.
        Left column shows reduced $\mime=200$, right column shows true
        proton-electron $\mime=1836$.
        Top row: initial upstream distribution in $(v_\prll,v_\perp)$,
        normalized to $c_{se}$; this corresponds to $v_{\prll 0}$ and
        $v_{\perp 0}$ in Equation~\eqref{eq:vprll-evol}.
        Red line bounds the region wherein electrons will mirror reflect upon
        encountering the shock.
        Bottom row: downstream electron distribution in $(v_\prll,v_\perp)$.
        The colormap shows phase-space density in arbitrary units, but
        matched in all panels to allow quantitative comparison.
    }
    \label{fig:liouville-mime}
\end{figure}

Figure~\ref{fig:liouville-mime} summarizes the just-preceding discussion by
showing the test-particle
electron Liouville mapping for two idealized shocks having
identical $\Ms$, $\tilde{\phi}_\prll(x)$, and $B(x)$ and differing only in mass
ratio $\mime$; see, e.g., \citet{yuan2008} for a more sophisticated example of
such test-particle mapping.
The Liouville map is performed following
Equations~\eqref{eq:liouville-map-perp} to \eqref{eq:dxdx0}.
For a fixed starting point $(v_{\perp 0},v_{\prll 0})$, the Liouville mapping
(phase space flow) from pre- to post-shock is identical and has no dependence
upon mass ratio, when specified in terms of upstream electron thermal velocity
$c_{se}$.
But, the starting electron distribution is offset farther from $v_{\prll 0}=0$
in the $\mime=200$ case; this leads to a stronger $v_\prll<0$ beam and a more
asymmetric post-shock distribution in the HT frame, as compared to the
$\mime=1836$ case.

It is not a good assumption that $\tilde{\phi}_\prll$ is the same for two
shocks of different $\mime$.
Indeed, $\tilde{\phi}_\prll$ decreases with $\mime$ in our 1D $\Ms=4$ example
by a factor of $2$ between $\mime=200$ to $400$, and then appears unchanged
for higher $\mime$.
Nevertheless, as we have just shown, it is helpful to separate the effects of
$u_{HT}/c_{se}$ and $\tilde{\phi}_\prll$ upon electron energization.

\begin{figure*}
    \includegraphics[width=7in]{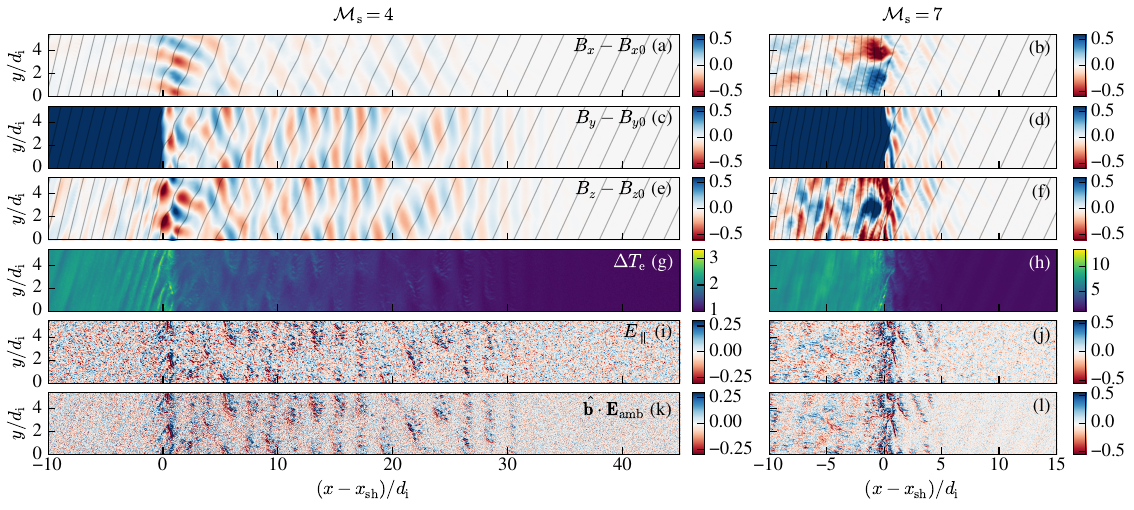}
    \caption{
        2D structure of $\Ms=4$ (left column) and $\Ms=7$ (right column),
        $\thetaBn=65^\circ$ shocks.
        Note that color scaling is different for each shock (column).
        (a--b): Magnetic fluctuation $(B_x-B_{x0})/B_0$.
        (c--d): Magnetic fluctuation $(B_y-B_{y0})/B_0$.
        (e--f): Magnetic fluctuation $(B_z-B_{z0})/B_0$.
        (g--h): Electron temperature change $\Delta \Te = \Te/T_0 - 1$.
        (i--j): Parallel electric field $E_\prll$.
        (k--l): Parallel ambipolar electric field
        $\uvec{b}\cdot\vec{E}_\mathrm{amb}$.
        Black contours in (a--f) show magnetic field lines.
    }
    \label{fig:flds2d}
\end{figure*}

\begin{figure}
    \includegraphics[width=3.375in]{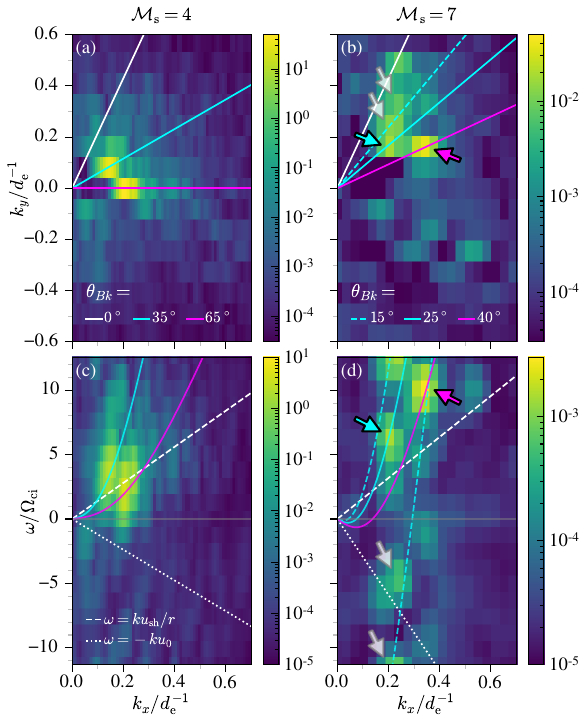}
    \caption{
        Fourier power spectra of $B_z$ in 2D $\Ms=4$ (left column) and 2D $\Ms=7$
        (right column) shock precursors show evidence for forward-propagating
        whistlers oblique to both $\uvec{n}$ and $\vec{B}$.
        (a): Power spectrum in ($k_x$,$k_y$) for $\Ms=4$.
        Lines mark angles $\thetaBk = 0^\circ$ (white), $35^\circ$ (cyan),
        and $65^\circ$ (magenta); $\thetaBk$ is the angle between
        wavevector $\vec{k}$ and the upstream magnetic field at
        $\thetaBn=65^\circ$.
        (b): Like panel (a) for $\Ms=7$.
        Lines mark $\thetaBk = 0^\circ$ (white), $15^\circ$ (dashed cyan),
        $25^\circ$ (cyan), and $40^\circ$ (magenta).
        Whistler power at $\thetaBk=25^\circ$ and $40^\circ$ is marked by cyan
        and magenta arrows; power at smaller $\thetaBk \lesssim15^\circ$ is
        shown by translucent gray arrows.
        (c): Power spectrum in ($k_x$,$\omega$) for $\Ms=4$.
        Solid lines are cold whistler dispersion Equation~\eqref{eq:whistler},
        Doppler-shifted from upstream plasma frame into simulation frame;
        colors correspond to angles marked in (a).
        Relevant flow speeds are shown as $\omega = k_x (u_\mt{sh}/r)$ (dashed
        white) and $\omega = - k_x u_0$ (dotted white).
        (d): Like panel (c) for $\Ms=7$.
        Frequency aliasing for $\thetaBk=15^\circ$ whistler dispersion is shown
        by cyan dashed curve wrapping around the top and bottom panel edges.
        Arrows in (d) mark our interpretation of the frequency structure
        corresponding to same arrows in (b).
    }
    \label{fig:fft}
\end{figure}

\begin{figure}
    \includegraphics[width=3.375in]{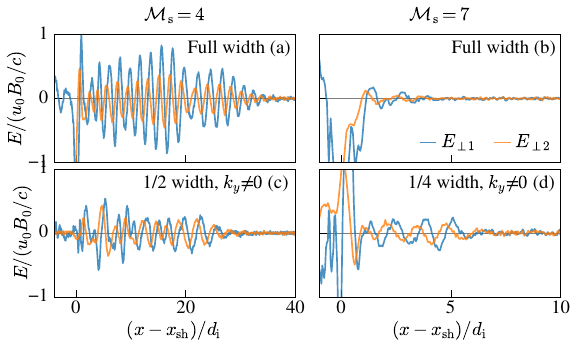}
    \caption{
        Shock precursor electric fields show right-handed polarization
        for both 2D $\Ms=4$ (left column) and 2D $\Ms=7$ (right column), when
        boosted into the upstream rest frame and projected into
        $\vec{B}_0$-perpendicular components.
        Right-handedness is shown by $E_{\perp 1}$ component offset in $x$ one
        quarter cycle ahead of $E_{\perp 2}$ component.
        (a--b): Electric fluctuations $E_{\perp 1}$ and $E_{\perp 2}$,
        $y$-averaged (i.e., only $k_y = 0$ mode) and scaled to the upstream
        motional field strength $u_0 B_0/c$ in the simulation frame.
        (c--d): Like (a--b), but show modes oblique to shock normal by
        computing $y$ average for one-half (c) or one-fourth (d) of the
        simulation domain, using Fourier-filtered fields with $k_y=0$ component
        forced to zero.
    }
    \label{fig:pol}
\end{figure}

\subsection{Precursor Wave Properties}
\label{sec:case-study-flds}

In this Subsection, we verify that the precursor waves in our 2D $\Ms=4$ and
$\Ms=7$ shocks are upstream-propagating
whistler waves, oblique to both $\vec{B}$ and $\uvec{n}$, with phase and group
speeds exceeding the shock speed.
It is already well established that weak, low-beta, quasi-perpendicular solar
wind shocks are formed of, regulated by, and radiate low frequency
($\omega\ll\Omce$) whistler-branch modes; evidence is given by
theory \citep{bickerton1971,tidman1971,krasnoselskikh2002},
experiments \citep{robson1969},
observations \citep{fairfield1975,greenstadt1975,mellott1984,oka2006,hull2012,wilson2012,wilson2016-waves,wilson2017,oka2019,hull2020,lalti2022-whistler},
and fully-kinetic simulations \citep{liewer1991,savoini1994,savoini2010,riquelme2011}.
The goal is to frame our case-study shocks within a broader context.

Figure~\ref{fig:flds2d} shows the 2D structure of our case study shocks.
Both $\Ms=4$ and $7$ cases show electromagnetic precursors oblique to $\vec{B}$
(Figure~\ref{fig:flds2d}(a--f)).
The electron temperature map shows hot filaments
at $x-x_\mt{sh}=0$
in the $\Ms=4$ case
(Figure~\ref{fig:flds2d}(g--h)),
some of which appear connected to
strong bipolar electrostatic structures at $y=2\di$ and $5\di$ within the shock
ramp (Figure~\ref{fig:flds2d}(i--l)).
Small-scale ($<\di$) electrostatic structures appear in $E_\prll$, or
equivalently in $\uvec{b}\cdot\vec{E}_\mt{amb}$, and they inhabit wave troughs
of local $B$ minima within the shock precursor region $x > x_\mt{sh}$
(Figure~\ref{fig:flds2d}(i--l));
many such structures are bipolar electron holes (positive electric potential).

Figure~\ref{fig:fft} presents Fourier power spectra of $B_z$ within the
shock precursor regions in order to identify the ion-scale precursor wave modes
and to measure their propagation angles and phase speeds.
We choose $B_z$ to include all electromagnetic modes in the 2D simulation
domain.
We sample $B_z$ at $0.25\Omci^{-1}$ intervals for a duration $5\Omci^{-1}$
towards each simulation's end.
The spatial grid of $B_z$ used for analysis is $6\times$ downsampled in all
directions.
The spatial windows are $x-x_\mt{sh} = 2$--$45\di$ and $2$--$15\di$ in the
$\Ms=4$ and $\Ms=7$ cases respectively, at end of each simulation.
The spatial window is stationary in the simulation frame; at earlier time
snapshots, the window is farther from the shock, and the precursor gradually
advances into the spatial window over $5 \Omci^{-1}$.
Before computing the Fourier transform, we apply a Blackman-Harris window to
both $x$ and $t$ coordinates to reduce spectral leakage.
The frequencies and wavenumbers in Figure~\ref{fig:fft} are presented in
the simulation frame,
and $\thetaBk$ is the angle between wavevector $\vec{k}$ and the upstream
magnetic field at $\thetaBn=65^\circ$.
The plotted frequency range is mildly asymmetric because we use an even number of time
snapshots, so Nyquist-frequency power is assigned to the $\omega=+12.6\Omci^{-1}$ bin.

In Figure~\ref{fig:fft}(c--d),
we plot an approximate cold whistler dispersion relation from
\citet{krasnoselskikh1985,krasnoselskikh2002}:
\begin{equation}
    \label{eq:whistler}
    \frac{\omega^2}{\Omce^2}
    = \frac{1}{1 + (k\de)^{-2}}
    \left(
        \frac{\me}{\mi}
        + \frac{\cos^2\thetaBk}{1 + (k\de)^{-2}}
    \right) \, ,
\end{equation}
with $\omega$ in the plasma rest frame.
We plot Equation~\eqref{eq:whistler} with a Doppler shift into the simulation
frame, $\omega \to \omega - k_x u_0$; the simulation frame's upstream flow is
also shown as $\omega=-k_x u_0$ (dotted white line).
Due to our coarse time sampling, wave power can alias in frequency $\omega$,
and we account for this by also aliasing one of the dispersion curves
(Figure~\ref{fig:fft}(d), cyan dashed).

Wave phase velocities may be computed from Figure~\ref{fig:fft} as:
\begin{equation}\label{eq:phasevelconv}
    \frac{\omega}{kc}
    = \frac{\omega}{\Omci} \frac{1}{k_x\de} \frac{\vA}{c}
    \sqrt{\frac{\mi}{\me}} \cos\thetakn
    \, ,
\end{equation}
with upstream $\vA/c = 0.028$ and $0.020$ in our $\Ms=4$ and $\Ms=7$ shocks
respectively; $\thetakn$ is the angle between between $\vec{k}$ and shock
normal $\uvec{n} = \uvec{x}$.
Equation~\eqref{eq:phasevelconv} is only a unit conversion and vector
projection and does not use Equation~\eqref{eq:whistler}.

The $\Ms=4$ shock shows two coherent precursor wave trains
(Figure~\ref{fig:fft}(a),(c)).
The strongest precursor travels along the shock normal $\uvec{n}$; a second
precursor travels at an oblique angle $\abt 35^\circ$ with respect to both
$\vec{B}$ and $\uvec{n}$.
Both wave trains have projected phase speed $\omega/k_x$ near or above the
shock speed
$\omega/k_x = u_\mt{sh}/r = 0.028c$ (Figure~\ref{fig:fft}(c), dashed white line).
The shock normal-aligned train has $\omega=2.5\Omci$, $k_x=0.21\de^{-1}$, and
phase speed $\omega/k = \omega/k_x = 0.024c$.
The oblique wave train has $\omega=3.7\Omci$, $k_x=0.14\de^{-1}$, and phase
speed $\omega/k = (\omega/k_x)\cos\thetakn = 0.046c$.
We take $\omega$, $k_x$, and $k_y$ to be the location of 1D power spectrum
maxima within slices of the 2D $(\omega,k_x)$ and $(k_x,k_y)$ spectra.
Measurement uncertainty comes from the coarsest Fourier transform bins,
$\Delta k_y = 0.08\de^{-1}$ and $\Delta \omega = 1.3\Omci^{-1}$, so our phase
speed estimates are only good to tens of percent.
Nevertheless, we conclude that the shock normal-aligned train's
$\omega/k \approx u_\mt{sh}/r$
permits it to phase stand in the shock frame.
The oblique train's phase-velocity vector has shock-normal
($\uvec{n}=\uvec{x}$) projected component equal to
$(\omega/k)\cos\thetakn = (\omega/k_x)\cos^2\thetakn = 0.040c \gtrsim
u_\mt{sh}/r$, permitting it to co-move with or out-run the shock.

The $\Ms=7$ shock shows multiple modes that we attribute to forward-propagating
whistlers with propagation angles $\thetaBk \sim 0$--$40^\circ$.
In the $k_x$--$k_y$ spectrum (Figure~\ref{fig:fft}(b)), two low-$k_y$
modes appear at $\thetaBk=25^\circ$ and $40^\circ$ (cyan, magenta
arrows).
Wave power also appears at smaller $\thetaBk$ (gray arrows).
The wave power resolves into more distinct modes in $\omega$--$k_x$ space
(Figure~\ref{fig:fft}(b)); we verify that the $\thetaBk=25^\circ$ and
$40^\circ$ modes lie on the oblique whistler dispersion relation.
We also see weaker, distinct blobs of wave power within
$k_x \sim 0.1$--$0.4\de^{-1}$, at all frequencies within the Nyquist-limited
band $-11.3$ to $12.6\Omci$.
We interpret these blobs as frequency-aliased whistler modes at near-parallel
$\thetaBk \lesssim 15^\circ$ propagation angles (Figure~\ref{fig:fft}(d)).
The $k_x$--$k_y$ power
spectrum in conjunction with Equation~\eqref{eq:whistler} helps break the
frequency-aliasing degeneracy and supports our identification of the
waves as whistlers.
As an example, the aliased dispersion for $\thetaBk=15^\circ$ (dashed cyan)
crosses two blobs of negative frequency (gray arrows) in
Figure~\ref{fig:fft}(d), which we attribute to $k_x$ and $k_y$ of the
same $\thetaBk$ in the $k_x$--$k_y$ spectrum (Figure~\ref{fig:fft}(b), gray
arrows).
The angle $\thetaBk$ is uncertain; the same power could be explained by, e.g.,
doubly-aliased $\thetaBk = 0^\circ$ whistlers.
We cannot cleanly identify frequencies and wavevector angles due to both
aliasing and the coarse frequency- and wavenumber-space resolution.
But, we can conclude that the supercritical $\Ms=7$ shock hosts oblique
($\thetaBk \lesssim40^\circ$) whistler modes with simulation-frame frequencies
$\omega\gtrsim5\Omci$ around and above the lower hybrid range.
For the $\thetaBk=25^\circ$ and $40^\circ$ modes, we estimate that their shock
normal-projected phase velocities
$(\omega/k)\cos\thetakn = (\omega/k_x)\cos^2\thetakn \approx 0.026c$ and
$0.036c$ respectively, at or above the shock speed $u_\mt{sh}/r = 0.023c$ in
the simulation frame.

Figure~\ref{fig:pol} shows that the precursor waves are right-hand polarized,
for both shock-normal aligned and oblique modes, in both the $\Ms=4$ and
$\Ms=7$ 2D shocks.
Following \citet[Ch.~1]{stix1992}, polarization is defined by the rotation
sense of a wave's electric field about its background (upstream) magnetic
field $\vec{B}_0$ at a fixed point in space; counter-clockwise rotation
about $\vec{B}_0$ is right-handed polarization.
We project $\vec{E}$ along the right-handed coordinate triad of unit vectors
$(\uvec{n}_{\perp 1}, \uvec{n}_{\perp 2}, \uvec{b}_0)$, where
$\uvec{b}_0$ is the upstream magnetic field direction,
$\uvec{n}_{\perp 1} = \uvec{z}$, and
$\uvec{n}_{\perp 2} \propto \uvec{z}\times\uvec{b}_0$.\footnote{
    This coordinate system matches that of Figure~\ref{fig:vdf-salami} in the
    far upstream, where $\uvec{b}=\uvec{b}_0$ and $\uvec{V}_\mt{e} = -\uvec{n}$.
}
All precursor wavevectors lie at an acute angle with respect to $\uvec{x}$
(Figure~\ref{fig:fft}(a),(c)), so wave fronts advance toward $+\uvec{x}$.
Therefore, at a fixed time, right-handedness is shown by $E_{\perp 1}$
offset one quarter cycle ahead of $E_{\perp 2}$ when plotted as a function
of $x$ in Figure~\ref{fig:pol}.

We separate the $\uvec{n}$-parallel and $\uvec{n}$-oblique modes in order
to check their polarizations.
The $\uvec{n}$-parallel modes are shown in Figure~\ref{fig:pol}(a--b) by
averaging along $y$ to capture only the $k_y=0$ mode.
The $\uvec{n}$-oblique modes are shown in Figure~\ref{fig:pol}(c--d) by
Fourier transforming $\vec{E}$, setting its $k_y=0$ Fourier coefficient to
zero, undoing the transform, and then averaging either the bottom one-half
or one-fourth of the simulation domain along $y$.
The choice of one-half and one-fourth allows us to capture wave power at,
respectively, $k_y=0.082\de^{-1}$ for $\Ms=4$ (Fig.~\ref{fig:fft}(a)) and
$k_y=0.164\de^{-1}$ for $\Ms=7$ (Fig.~\ref{fig:fft}(b)); these modes have
either one or two standing wavelengths along $y$.
In the $\Ms=4$ case, we see that the $\uvec{n}$-parallel precursor wave has
higher amplitude and longer $x$ extent than the $\uvec{n}$-oblique wave;
in the $\Ms=7$ case, the opposite holds.
All four wavetrains show $E_{\perp 1}$ one quarter cycle ahead of
$E_{\perp 2}$ and are therefore right-hand polarized, consistent with a
fast mode having $\vec{k}$ at acute angle to $\uvec{x}$ (i.e., forward
propagating with respect to $x$).

\subsection{Precursor Wave Interpretation}
\label{sec:case-study-flds2}

\begin{figure*}
    \centering
    \includegraphics[width=6in]{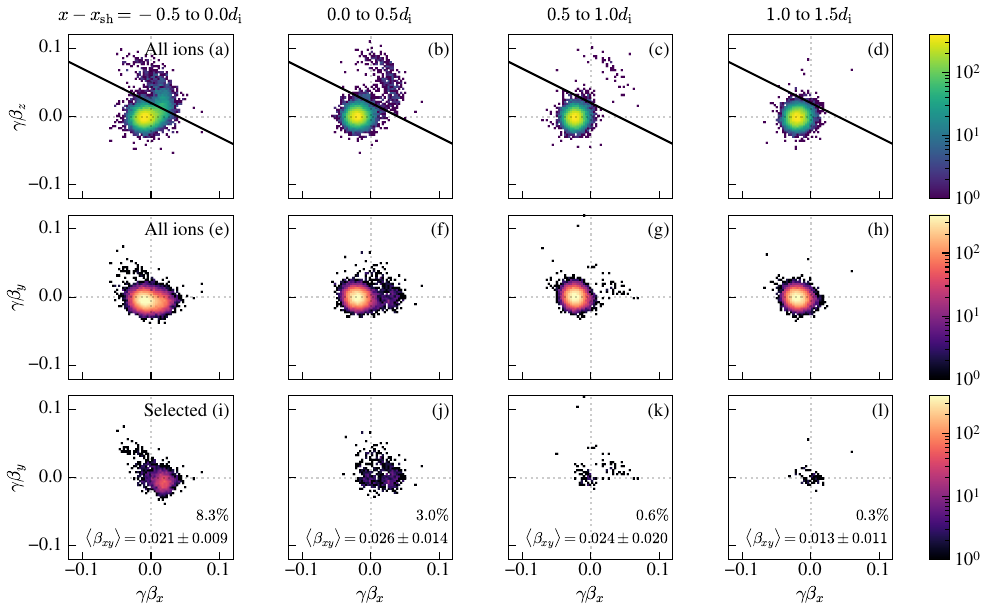}
    \caption{
        Ion momentum distributions near the shock front, 2D $\Ms=4$ case study,
        measured in $0.5\di$ wide bins along $x$ (left to right).
        Top row: $\gamma\beta_x$--$\gamma\beta_z$ distribution of all ions.
        Middle row: $\gamma\beta_x$--$\gamma\beta_y$ distribution of all ions.
        Bottom row: $\gamma\beta_x$--$\gamma\beta_y$ distribution of reflected
        ions, as chosen by cut in $\gamma\beta_x$--$\gamma\beta_z$ space (all
        ions above thick black line, top row).
        Bottom row also reports the reflected ion percentage (out of the total
        population) and the mean and standard deviation of the $x$-$y$ plane
        projected velocity $\beta_{xy}$, in units of $c$.
    }
    \label{fig:ion-vdf-Ms4}
\end{figure*}

\begin{figure*}
    \centering
    \includegraphics[width=6in]{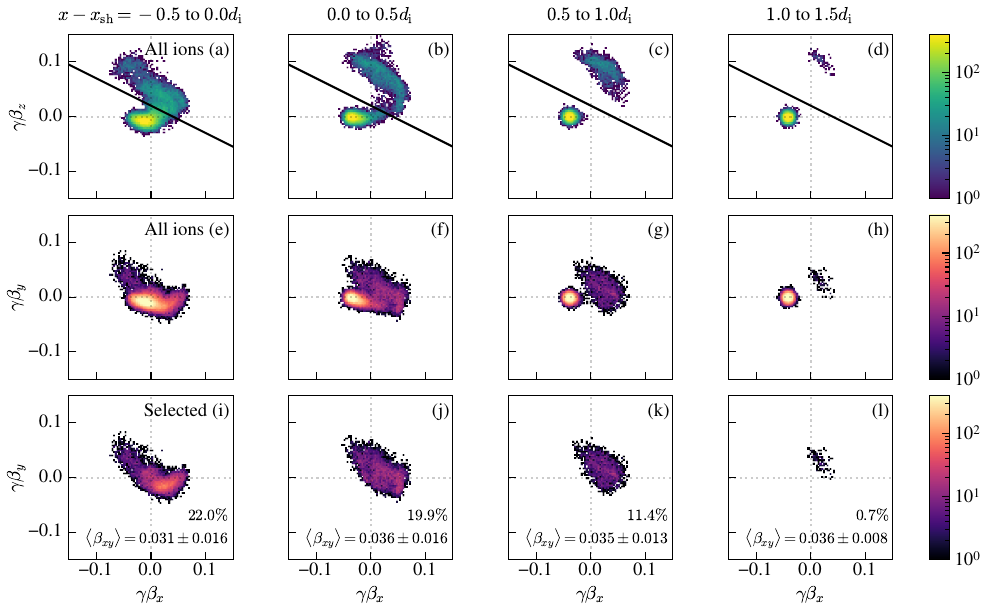}
    \caption{
        Like Figure~\ref{fig:ion-vdf-Ms4}, but for the 2D $\Ms=7$ case study.
    }
    \label{fig:ion-vdf-Ms7}
\end{figure*}

What is the origin of the precursor whistlers?
A full answer to this question is beyond the scope of our work, but we
shall give some comments.
In the $\Ms=4$ case, the
shock-normal whistlers are consistent with a phase-standing wave train expected
to form for $\Ma < \Mw$ \citep{kennel1985,krasnoselskikh2002}.
For the whistlers oblique to $\uvec{n}$, we consider two plausible
mechanisms:
nonlinear wave steepening within the shock ramp \citep{krasnoselskikh2002}, or
beam resonance with reflected ions gyrating within the shock foot, which may be
called ``modified two stream instability'' (MTSI) or ``kinetic cross-field
streaming instability''
\citep{wu1984,hellinger1996,hellinger1997,muschietti2017}.
There also exists a less-studied wave-wave decay instability that may generate
daughter whistlers having $\vec{k}$ mis-aligned with respect to their parent
whistlers \citep{galeev1963,decker1972}, noted by
\citet[pg.~2158]{mellott1984}, but we do not consider it
for now.

Lower hybrid drift instabilities have been invoked to explain shock structure
and electron heating \citep{krall1971,wu1984,stasiewicz2020-ii}; but, drift
waves propagate orthogonal to both $\vec{B}$ and $\del P_\mt{e}$ (i.e.,
$\uvec{n}$) and are thus suppressed in our 2D simulations.
We also do not consider ``slow'' MTSI caused by a relative drift between incoming
ions and electrons \citep{muschietti2017}.
Both lower hybrid drifts and ``slow'' MTSI are expected to drive waves nearly
perpendicular to $\vec{B}$ that are mainly electrostatic and so would not
explain the oblique, electromagnetic power that we are now considering.

Both $\Ms=4$ and $7$ shocks have $\Ma$ below the non-linear whistler Mach
number $\sqrt{2} \Mw$ as defined by \citet{krasnoselskikh2002}, and
Figure~\ref{fig:fft} shows that whistlers in range of $\thetaBk$ can
outrun both shocks.
Of note, our $\Ms=7$ shock having $\Mw < \Ma < \sqrt{2}\Mw$ is similar to the
Cluster shock crossing presented by \citet{dimmock2019}, wherein $\de$-scale
shock ramp fluctuation was presented as evidence for nonlinear steepening as
the origin of oblique whistlers.
More broadly speaking, it seems plausible that 2D shock ramp rippling or
filamentation, coupled with wave steepening in the ramp, may cause whistler
wave emission at various angles \citep[cf.][]{sundkvist2012}.

Can reflected ions resonate with the oblique precursor waves?
We assess this by comparing the reflected ions' $x$-$y$ plane velocity
$\beta_{xy} = v_{xy}/c$ against the precursor whistler phase speeds
(Equation~\eqref{eq:phasevelconv}), all measured in the simulation frame.
We separate reflected ions from ``core'' ions using an $x$-$z$ momentum-space
threshold $\gamma\beta_z > 0.02 - 0.5 \gamma\beta_x$, which selects ions with
high $v_x$ and $v_z$ being accelerated by the upstream motional field near the
shock ramp.
The same momentum-space cut is applied for both $\Ms=4$ and $\Ms=7$.
We compute the mean and standard deviation of $\beta_{xy}$ within $0.5\di$ bins
along $x$, noting that the reflected ion distributions are not Gaussian or
particularly symmetric in phase space.

Figures~\ref{fig:ion-vdf-Ms4} and \ref{fig:ion-vdf-Ms7} show the ion
distributions and measurements of $\beta_{xy}$.
Reflected ions appear within $\pm 1\di$ of $x_\mt{sh}$ for both the
$\Ms=4$ and $7$ shocks, with a shorter shock-foot gyration ($\lesssim 0.5\di$
along $x$) for the $\Ms=4$ shock.
In the last simulation snapshot, the reflected ions have a number density
contrast $\sim 3$--$8\%$ ($\Ms=4$) and $\abt 20\%$ ($\Ms=7$) with respect to
the total ion population.

The $\Ms=4$ shock's reflected ion beam has mean $v_{xy} \abt 0.02$--$0.026c$
(standard deviation $0.01c$).
It appears difficult for the reflected ions to resonate with the
$\thetaBk=35^\circ$ whistler mode at $\omega/k \sim 0.046c$, but the ions
could (in principle) lie in beam resonance with the shock-normal
$\thetaBk=65^\circ$ whistler at $\omega/k \sim 0.024c$.

The $\Ms=7$ shock's reflected ion beam has mean $v_{xy} \abt 0.03$--$0.04c$
(standard deviation $0.02c$).
The reflected ions might resonate with the lower-frequency $\thetaBk=40^\circ$
and $25^\circ$ whistler modes having $\omega/k \sim 0.033$ and $0.04c$
respectively.
The reflected ions may be too slow to resonate with the higher-frequency
$\thetaBk\lesssim 15^\circ$ whistlers, which have $\omega/k \gtrsim 0.05c$.

Our analysis is limited because we do not consider any of (1) time or spatial
(2D) variation in the reflected ion properties, (2) angle between reflected
ions' velocity vector and the precursor wavevectors, (3) energy or power
balance, (4) more detailed kinetic stability analysis.
A proper treatment of (1) may loosen our constraints if time-intermittent or
localized, gyrophase-bunched ion reflection can attain a larger range of beam
speeds and higher density with respect to the incoming flow.
A proper accounting of (2--4) should help tighten our constraints by culling
observed wave modes that could not plausibly be driven by a resonant ion beam.
Now, (1--4) are not unusually difficult to assess, but we emphasize that the
purpose of Sections~\ref{sec:case-study-flds} and \ref{sec:case-study-flds2} is
to provide broader context for our primary focus of electron heating physics.
We can at least conclude that reflected ions' beam resonance likely does not
explain all of the oblique wave precursor power observed in our simulations,
and perhaps nonlinear shock ramp processes are needed.

We warn that the propagation angle $\thetaBk$ of our whistlers is subject to
both a mode selection effect from the simulation domain's periodic boundary in
$y$, as well as angle-dependent Landau and transit-time damping that may differ
between our simulated $\mime=200$ and the true $\mime=1836$.
We also warn that our angles $\thetaBk$ are by-eye estimates, rather than being
obtained by a formal fitting procedure; we deem this acceptable since neither
precise wave angle determination nor wave driving mechanism determination is
the primary purpose of this manuscript, and there is considerable systematic
uncertainty due to the frequency aliasing in our $\Ms=7$ shock analysis
anyways.

Our most robust conclusion is that forward-propagating precursor whistlers
appear at a variety of propagation angles between $\thetaBk = 0^\circ$ to
$\thetaBk = \thetaBn$ in our case study shocks.
These whistlers are important in setting the ion-scale structure of the
electron's parallel potential.
We expect such precursor whistlers to occur in a wide range of simulated
parameter space as fully-kinetic simulations catch up to decades of spacecraft
observations.

\section{Conclusions}

\subsection{Summary}

We have measured electrons' ambipolar $\vec{B}$-parallel potential $\phiprll$ in a survey
of 2D quasi-perpendicular PIC shocks spanning $\Ms=3$--$10$ ($\Ma \sim 1$--$5$)
and $\thetaBn=85$--$55^\circ$, with upstream total plasma beta $\bp=0.25$.
Different particle- and field-based measurements of $\phiprll$ agree in the
overall $\phiprll$ jump at the $\abt 10\%$ level, and we find that $\phiprll$
is most robustly estimated in our simulations from the electron pressure tensor
divergence.
Off-diagonal terms of the electron pressure tensor $\vec{P}_\mt{e}$ are not
negligible for our low-$\bp$ shock simulations; the integral of the ambipolar
electric field $\vec{E}_\mt{amb}$ projected along shock normal can
underestimate $\phiprll$ by $\abt 10$--$30\%$ as compared to other measurement
methods.
Both the magnitude of $\phiprll$, and the correlation between $\phiprll$ and
$\Delta\Te$, are similar to prior reports based on observational data
\citep{schwartz1988,hull1998,hull2000-isee}.
We also measure the normal incidence frame (NIF) potential $\phinif$ and show
its variation with $\thetaBn$ and Mach number.

We show that a quasi-perpendicular shock with Alfv\'{e}n Mach number $\Ma$
below a critical whistler Mach number, $\Ma < \Mw$, can host a $\phiprll(x)$
profile extending over many tens of $\di$ within a whistler wave precursor.
The potential shows transient spikes within magnetic troughs, as well as a
secular increase towards the shock over many wave cycles.
We speculate that the potential spikes (bipolar electric fields) are due to
non-linear steepening of the large-amplitude whistler wave's electrostatic
field \citep{vasko2018-whistler,an2019}, but more work is needed to properly
explore and test this hypothesis.

In two case-study shocks ($\mime=200$, $\Ms=4$ and $7$, $\thetaBn=65^\circ$)
without backstreaming ions, we find that shock precursor whistlers span various
propagation angles between $\thetaBk=0^\circ$ to $\thetaBk=\thetaBn$, including
a phase-standing wave train along shock-normal in the $\Ms=4$ case
\citep{tidman1968,kennel1985}.
The whistler precursors also host small-scale electrostatic structures,
including electron holes, with wavevector $\vec{k}$ parallel to $\vec{B}$.
We tentatively find that shock-reflected (gyrating, not backstreaming) ions may
lie in beam resonance with some, but not all, of the precursor wave power.
Other mechanisms, such as nonlinear steepening within the shock ramp
\citep{omidi1988,omidi1990,krasnoselskikh2002} or wave-wave decay instability
\citep{galeev1963,decker1972}, may help explain some precursor wave power that
cannot be attributed to beam ions.

\subsection{Future Directions}

Our description of electron energization has been macroscopic,
focusing upon electric fields at ion scales;
we have not quantified the microscopic scattering and dissipation that is
needed to provide true irreversible heating and to regulate the strength of
$\phiprll$ via the electron pressure tensor.
Moreover, the case-study simulations of Section~\ref{sec:case-study} have
electron holes as the main electrostatic structure, whereas MMS observations
have revealed electron and ion holes, double layers, ion acoustic waves, and
electron cyclotron harmonics \citep{goodrich2018,wang2021}.
Towards higher $\bp$, small-scale parallel whistlers will likely become
important \citep{hull2012,page2021}.
Much recent work has been done with MMS and some tailored simulations
\citep{goodrich2018,vasko2018-esw,wang2021,shen2021,kamaletdinov2021,sun2022}.
Fewer fully-kinetic shock simulations have captured the macroscopic,
ion-scale coupling to microscopic electrostatic scales \citep{wilson2021}.
In such shock simulations, we would like to know: what types of electrostatic
structures appear, and where in the shock do said structures appear, as a
function of shock parameters $\Ms$, $\thetaBn$, and $\bp$?
What is their contribution to $\vec{B}$-parallel scattering and pitch-angle
scattering of electrons?
As the latter two questions are answered, can we find any qualitative or
quantitative trends in shock parameter space to test against MMS data?

A few caveats must be noted for further study of these secondary electrostatic
structures.
First, our simulations are limited by spatial resolution and numerical noise.
The upstream electron Debye length is marginally resolved with one cell, and
not resolved when PIC current filtering is accounted for; small-scale
structures may be biased towards lower $k$.
Discrete PIC macroparticles create numerical scattering \citep{birdsall1991}
whose contribution needs to be separated from that of more-physical
collisionless scattering.
High mass ratio $\mime$, high particle sampling, and high space and time
resolution may be needed to unveil physics at the electron Debye scale in
future simulations; Vlasov solvers may be another way to make progress
\citep[e.g.][]{juno2018}.
Second, we have only used equal-temperature ion-electron Maxwellian
distributions to model upstream plasma in our shocks.
Solar wind shocks have multiple populations: core, halo, and strahl electrons,
pick-up ions, et cetera.
The electron halo and strahl, in particular, may alter Landau damping rates.
Third, artificially low $\ompe/\Omce$ can modify the types and Fourier spectra
of electrostatic waves, e.g., for the electron cyclotron drift instability
(ECDI) \citep{muschietti2013,tenbarge2021-dpp}, and hence also modify the
energy exchange between ions and electrons.
Lastly, the use of 2D versus 3D simulations will matter -- just as the electron
trapping is weaker in 2D versus 1D, so we also expect that 3D simulations may
reduce electron trapping, admit a broader spectrum of precursor wave modes, and
permit electrons to scatter and gain energy in different and possibly new ways
\citep{trotta2019}.

\begin{acknowledgments}
We are grateful to anonymous referees whose comments have helped to stimulate
much of this study.
We acknowledge helpful conversations, brief or extended, with
Takanobu Amano,
Artem Bohdan,
Collin Brown,
Li-Jen Chen,
Luca Comisso,
Greg Howes,
Arthur Hull,
Jimmy Juno,
Matthew Kunz,
Steve Schwartz,
Navin Sridhar,
Jason TenBarge,
Ivan Vasko,
Lynn B.~Wilson III,
participants and organizers of the Geospace Environment Modeling (GEM) Focus
Group ``Particle Heating and Thermalization in Collisionless Shocks in the MMS
Era'', and many other members of the scientific community.
We especially thank Lynn for alerting us to a Liouville mapping
normalization error in an earlier version of this manuscript.
We are grateful for the use of analysis and simulation code developed by Xinyi
Guo.

AT was supported by NASA FINESST 80NSSC21K1383 and a NASA Cooperative Agreement
awarded to the New York Space Grant Consortium;
AT acknowledges travel support from Columbia's Arts \& Sciences Graduate
Council and Department of Astronomy.
AT and LS acknowledge partial support by NASA ATP 80NSSC20K0565 and NSF
AST-1716567.
LS acknowledges support from the Cottrell Scholars Award, and DoE Early Career
Award DE-SC0023015.
This research has benefited from the hospitality of the International Space
Science Institute (ISSI) in Bern through project \#520 ``Energy Partition
across Collisionless Shocks''.
The material contained in this document is based upon work supported by a
National Aeronautics and Space Administration (NASA) cooperative agreement. Any
opinions, findings, conclusions or recommendations expressed in this material
are those of the author and do not necessarily reflect the views of NASA.

Simulations and analysis used the computer clusters Habanero, Terremoto,
Ginsburg (Columbia University), and Pleiades allocation s2610 (NASA).
Computing resources were provided by Columbia University's Shared Research
Computing Facility (SRCF) and the NASA High-End Computing Program through the
NASA Advanced Supercomputing (NAS) Division at Ames Research Center.
Columbia University's SRCF is supported by NIH Research Facility Improvement
Grant 1G20RR030893-01 and the New York State Empire State Development, Division
of Science Technology and Innovation (NYSTAR) Contract C090171.
This work was expedited by NASA's Astrophysics Data System, Jonathan Sick's and
Rui Xue's \texttt{ads2bibdesk}, and Benty Fields.
\end{acknowledgments}

\appendix

\section{Post-Shock Potential and Temperature Measurements}
\label{app:deltapot-measure}

\begin{figure*}
    \includegraphics[width=\textwidth]{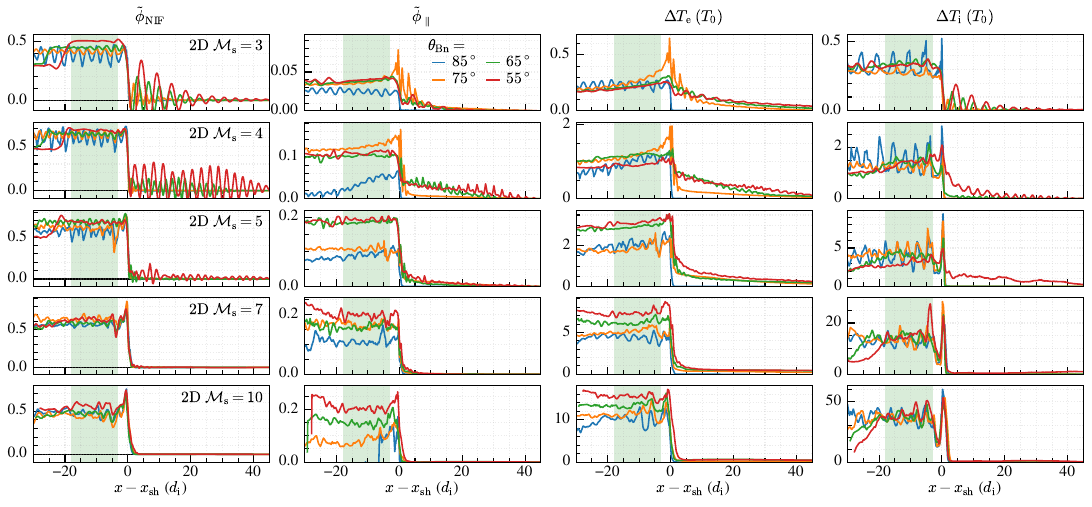}
    \includegraphics[width=\textwidth]{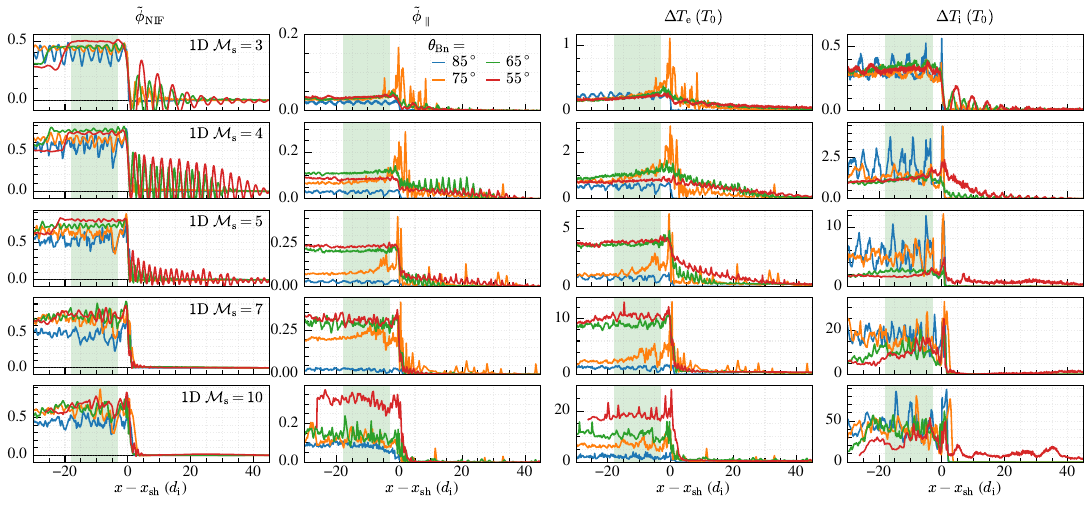}
    \caption{
        Measurement of $\Delta\phi_\mt{NIF}$,  $\Delta\phi_\parallel$,
        $\Delta T_\mathrm{e}$, and $\Delta T_\mathrm{i}$
        in simulated 2D and 1D shocks.
        Left column: normal incidence frame (NIF) potential
        $\phi_\mt{NIF} = \int E_x\dtl x$.
        Second from left column: parallel potential $\phi_\prll$, approximated
        by $\phiamb$ as described in Section~\ref{sec:phiprllres}.
        Third from left column: electron temperature jump $\Delta \Te$ scaled
        to upstream value $T_0$.
        Right column: ion temperature jump $\Delta\Ti$ scaled to upstream
        value $T_0$.
        Top five rows show 2D shocks, with $\Ms$ increasing from top to bottom;
        colors indicate $\thetaBn=85^\circ$ (blue), $\thetaBn=75^\circ$
        (orange), $\thetaBn=65^\circ$ (green), and $\thetaBn=55^\circ$ (red).
        Bottom five rows are organized like the top five rows, but for 1D shocks.
        We average $\phi_\mt{NIF}$, $\phi_\prll$, $\Te$, and $\Ti$ within the
        post-shock region $x-x_\mt{sh}=-18$ to $-3\di$ (light green vertical
        bands) to obtain the data plotted in Figures~\ref{fig:machscale},
        \ref{fig:deltapot-scatter-te}, and \ref{fig:deltapot-scatter-mach}.
    }
    \label{fig:deltapot-measure}
\end{figure*}

Figure~\ref{fig:deltapot-measure} shows the 1D, $y$-averaged profiles of
$\tilde{\phi}_\mt{NIF}$, $\tilde{\phi}_\prll$, $\Delta\Te$, and $\Delta\Ti$ for
our full set of 2D and 1D shocks.
All potentials and temperature profiles are computed using 50 finely-spaced
snapshots following the alignment and averaging procedure as described in
Section~\ref{sec:phiprllres}.
We compute the volume-averaged mean of $\phi_\prll$ and $\phi_\mt{NIF}$ to
obtain the single-point jumps $\Delta\phi_\prll$ and $\Delta\phi_\mt{NIF}$
plotted in
Figures~\ref{fig:deltapot-scatter-te}--\ref{fig:deltapot-scatter-mach}.
The mean electron and ion temperatures are weighted by their corresponding
species density.
We compute the mean $\phi_\prll$, $\phi_\mt{NIF}$, and $\Delta\Te$ in the
region $x-x_\mt{sh}=-18$ to $-3\di$, highlighted by light green in
Figure~\ref{fig:deltapot-measure}.
The normal-incidence frame (NIF) potential $\phi_\mt{NIF}$ is measured in
the simulation frame, since the boost to the NIF frame does not alter
$E_x$.

\section{Best-Fit Liouville-Mapped Distributions}
\label{app:liouville}

\begin{figure*}
    \includegraphics[width=\textwidth]{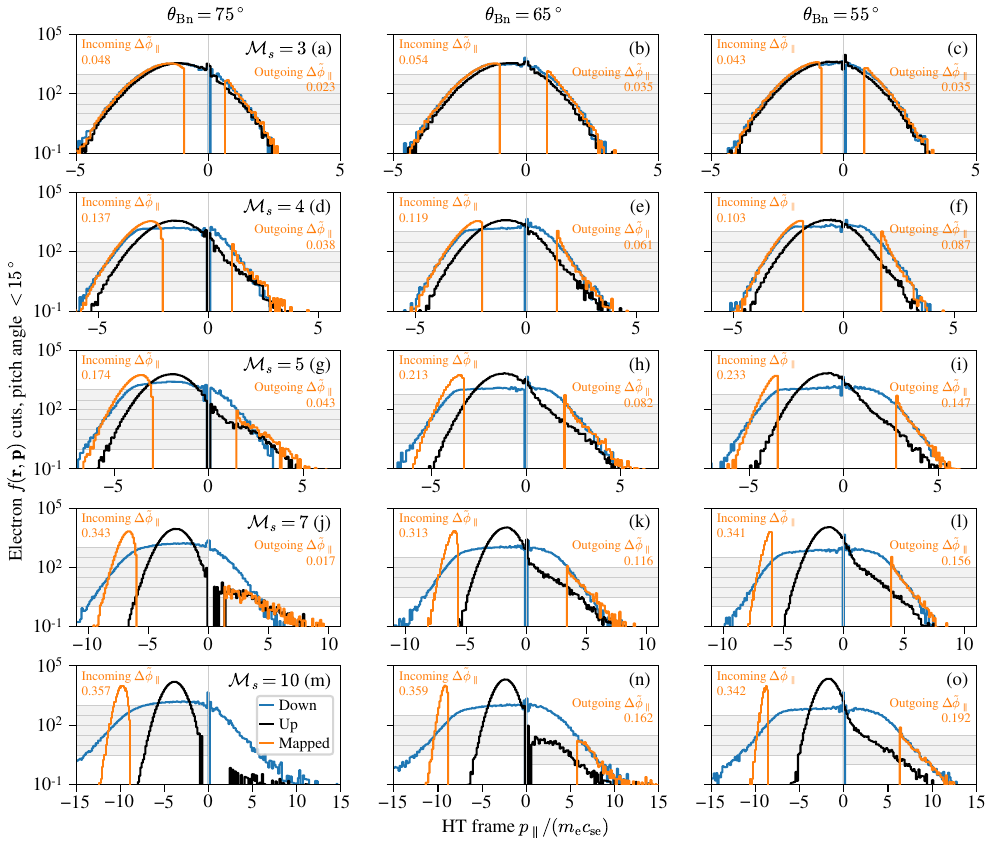}
    \caption{
        Like Figure~\ref{fig:liouville-fit-procedure}, but for multiple shocks
        in our 2D shock parameter sweep, and showing the Liouville-mapped
        distributions for the mean best-fit values of incoming and outgoing
        $\Delta\tilde{\phi}_\parallel$ instead of a range of best-fit mapped
        distributions.
        The mean values are also reported as red triangles in
        Figure~\ref{fig:dc-proxies-all}.
        In the figure legend, ``Up'', ``Down'', and ``Mapped'' respectively
        correspond to upstream, downstream, and Liouville-mapped
        distributions.
    }
    \label{fig:liouville-fit}
\end{figure*}

In Figure~\ref{fig:liouville-fit}, we show Liouville-mapped distributions
for many shocks in our parameter sweep.
Figure~\ref{fig:liouville-fit} is structured similarly to
Figure~\ref{fig:liouville-fit-procedure}, except that
it shows only one
Liouville-mapped distribution for
the mean best-fit $\Delta\tilde{\phi}_\prll$ value
on each side of $p_\prll = 0$.
We
observe that weaker, more oblique shocks (upper right of
Figure~\ref{fig:liouville-fit}) show more adiabatic electron behavior,
while stronger and more perpendicular shocks are not well described by
Liouville mapping for incoming ($p_\prll<0$) electrons.

\section{Electron Kinetic Energy in the HT Frame}

\begin{figure*}
    \includegraphics[width=\textwidth]{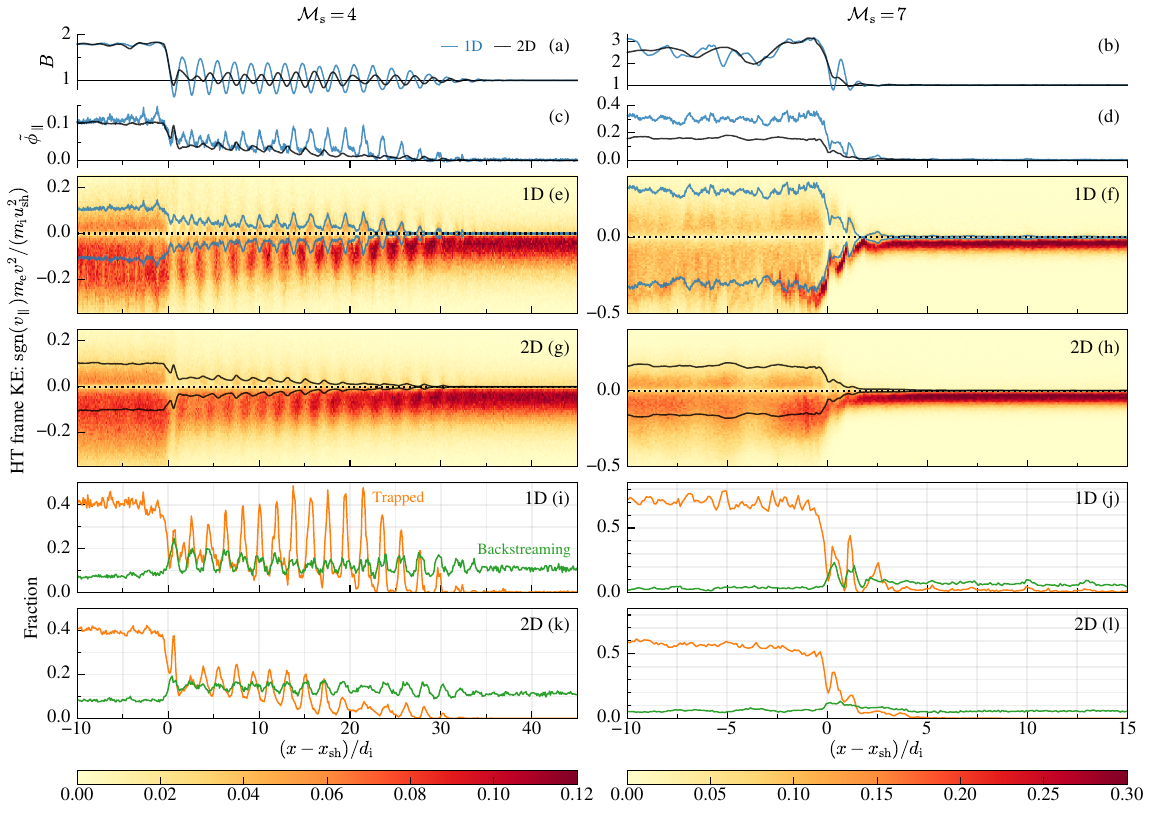}
    \caption{
        Like Figure~\ref{fig:xvprll}, except panels (e)--(h) are changed to
        show HT frame electron kinetic energy as a function of $x-x_\mt{sh}$.
        Electron energy is normalized to the ion bulk kinetic energy
        $\mi u_\mt{sh}^2/2$, like the overplotted parallel potential
        $\pm\tilde{\phi}_\prll(x)$.
        Electron energy is also multiplied by $\sgn(v_\prll)$ to show the
        direction of travel along $x$.
        All other panels are unchanged from Figure~\ref{fig:xvprll}.
    }
    \label{fig:xvprll-energy}
\end{figure*}

Figure~\ref{fig:xvprll-energy} shows electron kinetic energy, multiplied by
$\sgn(v_\prll)$, as an alternative to $x$-$v_\prll$ phase space in
Figure~\ref{fig:xvprll}.
The electron kinetic energy so plotted has a direct mapping to the ``Trapped''
and ``Backstreaming'' fractions in Figure~\ref{fig:xvprll-energy}(i)--(l);
recall that trapped electrons have $\varepsilon < 0$, and backstreaming
electrons have $\varepsilon > 0$ and $v_\prll > 0$.
In Figure~\ref{fig:xvprll-energy}(e)--(h), electrons with signed kinetic energy
between the $\pm \tilde{\phi}_\prll$ curves are trapped, while those above the
$+\tilde{\phi}_\prll$ curve are backstreaming.

\section{Numerical Convergence of Shock Structure} \label{app:ppc}

\begin{figure*}
    \centering
    \includegraphics[width=7in]{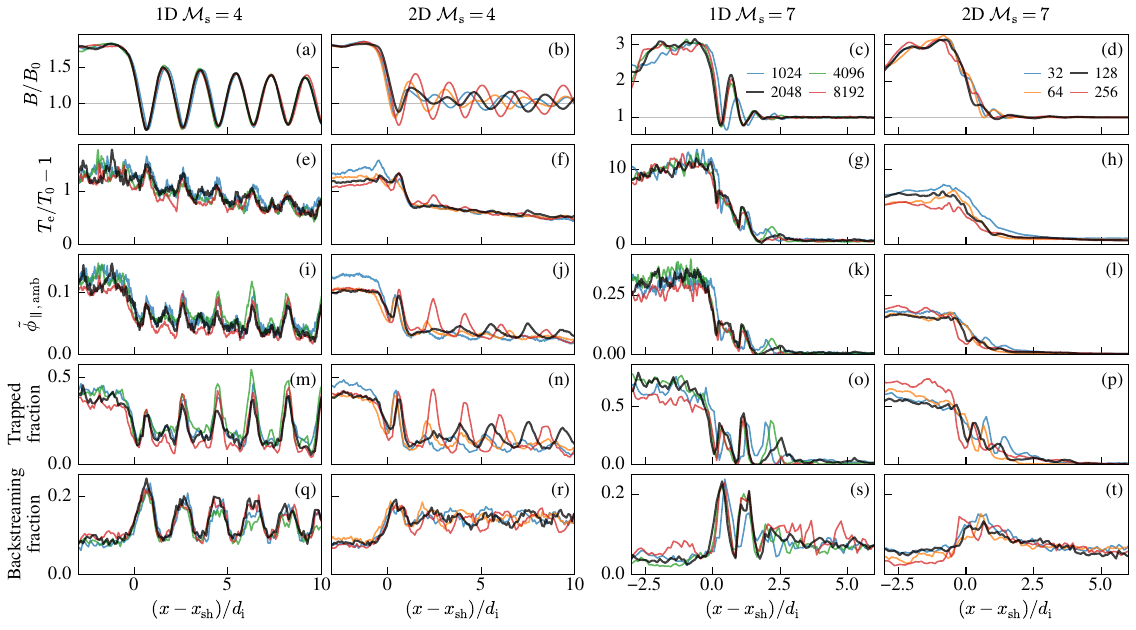}
    \caption{
        Particle per cell (PPC) convergence for four case-study simulations:
        1D $\Ms=4$ (left column), 2D $\Ms=4$ (second from left column),
        1D $\Ms=7$ (third from left column), and 2D $\Ms=7$ (right column).
        Upstream angle $\thetaBn=65^\circ$ for all shocks.
        In all panels, black curves are fiducial simulations with 2048 (1D) and
        128 PPC (2D).
        Colored curves are simulations with varying PPC: 1024--8192 (1D) and
        32--256 (2D).
        All quantities are $y$ averaged.
        (a--d): Magnetic field strength $B/B_0$.
        (e--h): Electron temperature jump $\Te/T_0 - 1$.
        (i--l): Parallel potential $\phiamb$, in units of $0.5 \mi u_\mt{sh}^2$.
        (m--p): Trapped electron fraction as defined for
        Figure~\ref{fig:xvprll}.
        (q--t): Backstreaming electron fraction as defined for
        Figure~\ref{fig:xvprll}.
    }
    \label{fig:conv}
\end{figure*}

We use our case study shocks to assess numerical convergence with respect to
the upstream number of particles per cell (PPC), both ions and electrons.
Recall that we adopted 2048 and 128 PPC for 1D and 2D simulations respectively.

The overall shock structure is well converged for all of our case studies
(Figure~\ref{fig:conv}(a)-(d)).
In 2D, the shock ramp and precursor waves are not matched in phase for
simulations with different PPC; the 1D simulations, constrained to a single
wave train along $x$, are more coherent.
The electron energy gain, as measured by both $\Te$ and $\phi_{\prll,\mt{amb}}$,
appears converged to within $10\%$ (Figure~\ref{fig:conv}(e)-(l)).
For the 2D $\Ms=7$ shock, the upstream to downstream jumps in $\Te$ and
$\phi_{\prll,\mt{amb}}$ do not vary monotonically with PPC, which we interpret
as statistical fluctuation rather than a lack of convergence.

The trapped and backstreaming electron fractions (Figure~\ref{fig:conv}(m)-(t))
also appear converged in all of our case study shocks.
The trapped fraction, in particular, serves as a coarse proxy for local phase
mixing and scattering of both $v_\prll$ and pitch angle within the whistler
precursor.
The electrostatic power in numerical noise is not small, being within an order
of magnitude of the power in electron holes and small-scale structures
(Figure~\ref{fig:flds2d}(i)--(l)).
It is not clear if the \emph{relative} contributions of different particle
scattering mechanisms are converged.
But, at least, Figure~\ref{fig:flds2d}(i)--(l) suggests that the
\emph{total} phase-space flow in our shocks is insensitive to PPC sampling.

\section{Simulation Parameters}

Table~\ref{tab:param} provides simulation input parameters and some
derived parameters for the main parameter sweep of 1D and 2D shocks, as well as
a set of 1D simulations with varying $\ompe/\Omce$ and $\mi/\me$ presented in
Section~\ref{sec:mime-dgamr}.
The parameters $\sigma$, $\delgame$, and $u_0$ are actual code inputs, from
which the shock parameters $\Ms$, $\Ma$, $\Mms$, $\bp$, $u_\mt{sh}$,
$u_\mt{sh}/r$ can be derived assuming a single-fluid adiabatic index
$\Gamma=5/3$.
The columns of Table~\ref{tab:param} not already defined in the main text are
as follows.
\begin{itemize}
    \item $t\Omci$ is the simulation duration.
    \item $\sigma = B_0^2 / \left[ 4\pi \left(\gamma_0 - 1\right) \left(\mi+\me\right) n_0 c^2 \right]$
        is a ratio of upstream magnetic and kinetic energy density,
        with $\gamma_0 = 1/\sqrt{1-u_0^2/c^2}$.
    \item $u_y$ is the post-shock transverse drift along $y$ predicted from the
        R-H conditions and used to set the electromagnetic fields at the
        simulation's left boundary.
    \item $u_\mt{right}$ is the manually-chosen expansion speed of the domain's
        right-side boundary.
\end{itemize}

\begin{deluxetable*}{lrrrrrrrrrrrr}
\tablecaption{
    Simulation input parameters.
    \label{tab:param}
}
\tablehead{
    \multicolumn{1}{l}{$\mi/\me$}
      & \multicolumn{1}{r}{$\Ms$}
      & \multicolumn{1}{r}{$\Ma$}
      & \multicolumn{1}{r}{$\Mms$}
      & \multicolumn{1}{r}{$\thetaBn$ (${}^\circ$)}
      & \multicolumn{1}{r}{$t \Omci$}
      & \multicolumn{1}{r}{$\sigma$}
      & \multicolumn{1}{r}{$\delgame$}
      & \multicolumn{1}{r}{$u_0/c$}
      & \multicolumn{1}{r}{$u_\mt{sh}/c$}
      & \multicolumn{1}{r}{$u_\mt{sh}/r/c$}
      & \multicolumn{1}{r}{$u_y/c$}
      & \multicolumn{1}{r}{$u_\mt{right}/c$}
}
\startdata
200 & 3 & 1.37 & 1.25 & 85 & 40.0 & 1.76E+1 & 1.01E$-$2 & 9.56E$-$3 & 3.89E$-$2 & 2.93E$-$2 & $-$5.89E$-$4 & 4.09E$-$2 \\
200 & 3 & 1.37 & 1.25 & 75 & 40.0 & 1.71E+1 & 1.01E$-$2 & 9.72E$-$3 & 3.89E$-$2 & 2.91E$-$2 & $-$1.81E$-$3 & 6.11E$-$2 \\
200 & 3 & 1.37 & 1.25 & 65 & 40.0 & 1.60E+1 & 1.01E$-$2 & 1.00E$-$2 & 3.89E$-$2 & 2.88E$-$2 & $-$3.17E$-$3 & 7.42E$-$2 \\
200 & 3 & 1.37 & 1.25 & 55 & 40.0 & 1.45E+1 & 1.01E$-$2 & 1.05E$-$2 & 3.89E$-$2 & 2.83E$-$2 & $-$4.76E$-$3 & 1.13E$-$1 \\
200 & 4 & 1.83 & 1.66 & 85 & 40.0 & 2.98E+0 & 1.01E$-$2 & 2.33E$-$2 & 5.18E$-$2 & 2.86E$-$2 & $-$1.10E$-$3 & 4.42E$-$2 \\
200 & 4 & 1.83 & 1.66 & 75 & 40.0 & 2.93E+0 & 1.01E$-$2 & 2.34E$-$2 & 5.18E$-$2 & 2.84E$-$2 & $-$3.33E$-$3 & 7.21E$-$2 \\
200 & 4 & 1.83 & 1.66 & 65 & 40.0 & 2.83E+0 & 1.01E$-$2 & 2.38E$-$2 & 5.18E$-$2 & 2.80E$-$2 & $-$5.62E$-$3 & 8.11E$-$2 \\
200 & 4 & 1.83 & 1.66 & 55 & 40.0 & 2.70E+0 & 1.01E$-$2 & 2.44E$-$2 & 5.18E$-$2 & 2.74E$-$2 & $-$8.00E$-$3 & 1.21E$-$1 \\
200 & 5 & 2.28 & 2.08 & 85 & 40.0 & 1.28E+0 & 7.08E$-$3 & 2.97E$-$2 & 5.42E$-$2 & 2.45E$-$2 & $-$1.10E$-$3 & 3.54E$-$2 \\
200 & 5 & 2.28 & 2.08 & 75 & 40.0 & 1.26E+0 & 7.08E$-$3 & 2.99E$-$2 & 5.42E$-$2 & 2.43E$-$2 & $-$3.29E$-$3 & 7.06E$-$2 \\
200 & 5 & 2.28 & 2.08 & 65 & 40.0 & 1.23E+0 & 7.08E$-$3 & 3.03E$-$2 & 5.42E$-$2 & 2.40E$-$2 & $-$5.45E$-$3 & 8.47E$-$2 \\
200 & 5 & 2.28 & 2.08 & 55 & 40.0 & 1.18E+0 & 7.08E$-$3 & 3.09E$-$2 & 5.42E$-$2 & 2.34E$-$2 & $-$7.56E$-$3 & 1.36E$-$1 \\
200 & 7 & 3.20 & 2.91 & 85 & 30.0 & 4.77E$-$1 & 5.06E$-$3 & 4.11E$-$2 & 6.41E$-$2 & 2.31E$-$2 & $-$9.71E$-$4 & 3.48E$-$2 \\
200 & 7 & 3.20 & 2.91 & 75 & 30.0 & 4.73E$-$1 & 5.06E$-$3 & 4.13E$-$2 & 6.41E$-$2 & 2.29E$-$2 & $-$2.88E$-$3 & 8.64E$-$2 \\
200 & 7 & 3.20 & 2.91 & 65 & 30.0 & 4.64E$-$1 & 5.06E$-$3 & 4.16E$-$2 & 6.41E$-$2 & 2.25E$-$2 & $-$4.67E$-$3 & 1.12E$-$1 \\
200 & 7 & 3.20 & 2.91 & 55 & 30.0 & 4.52E$-$1 & 5.06E$-$3 & 4.22E$-$2 & 6.41E$-$2 & 2.20E$-$2 & $-$6.25E$-$3 & 1.67E$-$1 \\
200 & 10 & 4.56 & 4.15 & 85 & 20.2 & 1.99E$-$1 & 3.04E$-$3 & 4.93E$-$2 & 7.09E$-$2 & 2.18E$-$2 & $-$6.69E$-$4 & 3.24E$-$2 \\
200 & 10 & 4.56 & 4.15 & 75 & 20.2 & 1.98E$-$1 & 3.04E$-$3 & 4.94E$-$2 & 7.09E$-$2 & 2.16E$-$2 & $-$1.96E$-$3 & 9.24E$-$2 \\
200 & 10 & 4.56 & 4.15 & 65 & 20.2 & 1.96E$-$1 & 3.04E$-$3 & 4.97E$-$2 & 7.10E$-$2 & 2.13E$-$2 & $-$3.12E$-$3 & 1.53E$-$1 \\
200 & 10 & 4.56 & 4.15 & 55 & 20.2 & 1.92E$-$1 & 3.04E$-$3 & 5.01E$-$2 & 7.09E$-$2 & 2.09E$-$2 & $-$4.04E$-$3 & 1.84E$-$1 \\
\hline
200 & 4 & 1.83 & 1.66 & 65 & 40.0 & 2.83E+0 & 2.02E$-$2 & 3.38E$-$2 & 7.33E$-$2 & 3.96E$-$2 & $-$7.95E$-$3 & 1.13E$-$1 \\
200 & 4 & 1.83 & 1.66 & 65 & 40.0 & 2.84E+0 & 5.06E$-$3 & 1.68E$-$2 & 3.66E$-$2 & 1.98E$-$2 & $-$3.97E$-$3 & 5.79E$-$2 \\
200 & 4 & 1.83 & 1.66 & 65 & 40.0 & 2.84E+0 & 2.53E$-$3 & 1.19E$-$2 & 2.59E$-$2 & 1.40E$-$2 & $-$2.81E$-$3 & 4.11E$-$2 \\
200 & 4 & 1.83 & 1.66 & 65 & 40.0 & 2.84E+0 & 1.26E$-$3 & 8.41E$-$3 & 1.83E$-$2 & 9.90E$-$3 & $-$1.98E$-$3 & 2.91E$-$2 \\
\hline
400 & 4 & 1.83 & 1.66 & 65 & 40.1 & 2.84E+0 & 1.01E$-$2 & 1.69E$-$2 & 3.67E$-$2 & 1.98E$-$2 & $-$3.98E$-$3 & 8.11E$-$2 \\
800 & 4 & 1.83 & 1.66 & 65 & 40.1 & 2.84E+0 & 1.01E$-$2 & 1.19E$-$2 & 2.60E$-$2 & 1.40E$-$2 & $-$2.81E$-$3 & 8.11E$-$2 \\
1836 & 4 & 1.83 & 1.66 & 65 & 40.1 & 2.84E+0 & 1.01E$-$2 & 7.87E$-$3 & 1.71E$-$2 & 9.27E$-$3 & $-$1.86E$-$3 & 8.11E$-$2
\enddata  
\tablecomments{
Table~\ref{tab:param} is
available in a machine-readable format in the online journal.
}
\end{deluxetable*}

\bibliographystyle{aasjournal}
\bibliography{library}

\end{document}